\documentclass[aps,pra,reprint,twocolumn,amsmath,amssymb,citeautoscript]{revtex4-1}
\usepackage{graphicx}
\usepackage{dcolumn}
\usepackage{bm}
\usepackage{latexsym,epsfig}
\usepackage{graphicx}
\usepackage[export]{adjustbox}
\usepackage{verbatim}
\usepackage{comment}
\usepackage{amsmath}
\usepackage{amssymb}
\usepackage{stmaryrd}
\usepackage{color}
\usepackage{epstopdf}
\usepackage{grffile}
\usepackage{mathtools}
\usepackage{physics}
\DeclareGraphicsExtensions{.eps}

\usepackage{float}
\usepackage{placeins}

\newcommand{\beq}{\begin{equation}}
\newcommand{\eeq}{\end{equation}}
\newcommand{\bea}{\begin{eqnarray}}
\newcommand{\eea}{\end{eqnarray}}
\newcommand{\ben}{\begin{eqnarray*}}
\newcommand{\een}{\end{eqnarray*}}
\newcommand{\bfig}{\begin{figure}}
\newcommand{\efig}{\end{figure}}

\usepackage{hyperref}
\hypersetup{
    colorlinks=true,      
    urlcolor=blue,
    citecolor=blue,
    linkcolor=blue
}

\definecolor{erc}{rgb}{1,0,1}

\begin{document}
\title{Phonon-induced breakdown of Thouless pumping in the Rice-Mele-Holstein model}
\author{Suman Mondal, Eric Bertok and Fabian Heidrich-Meisner}
\affiliation{Institut f\"{u}r Theoretische Physik, Georg-August-Universit\"{a}t G\"{o}ttingen, D-37077 G\"{o}ttingen, Germany}


\date{\today}

\begin{abstract}
Adiabatic and periodic variation of the lattice parameters can make it possible to transport charge through a system even without net external electric or magnetic fields, known as Thouless charge pumping. The amount of charge pumped in a cycle is quantized and entirely determined by the system's topology, which is robust against perturbations such as disorder and interactions. However, coupling to the environment may play a vital role in topological transport in many-body systems. We study the topological Thouless pumping, where the charge carriers interact with local optical phonons. The semi-classical multi-trajectory Ehrenfest method is employed to treat the phonon trajectories classically and charge carriers quantum mechanically. We find a breakdown of the quantized charge transport in the presence of phonons. It happens for any finite electron-phonon coupling strength at the resonance condition when the pumping frequency matches the phonon frequency, and it takes finite phonon coupling strength away from the resonance. Moreover, there exist parameter regimes with non-quantized negative and positive charge transport. The modified effective pumping path due to electron-phonon coupling accurately explains the underlying physics. In the large coupling regime where the pumping disappears, the phonons are found to eliminate the staggering of the onsite potentials, which is necessary for the pumping protocol. Finally, we present a stability diagram of quantized pumping as a function of time period of pumping and phonon coupling strength.
\end{abstract}

\maketitle

\section{Introduction}
A Thouless charge pump, on a fundamental level, is a one-dimensional dynamic equivalent to the quantum Hall effect. The robust transport of quantized charge in an adiabatic pump cycle, defined by periodic variation of lattice parameters, was first introduced almost four decades ago by Thouless~\cite{Thouless1983}. Recently, this fundamental physical phenomenon has been experimentally demonstrated in a variety of quantum systems, such as ultracold atoms in optical lattices and photonic lattices~\cite{Lohse2016,Schweizer2016, Takahashi2016pumping,Lahini2012,Yongguan2016,Alexander2020, Zlata2020}. The robustness of quantization is attributed to the underlying topological protection, which is not altered under small perturbations such as disorder and interaction. In recent years, studying the effect of disorder, different kinds of interaction and {non-linearity} has been a topic of immense interest~\cite{Titum2016,Ke2017,Hayward2018,Nakagawa2018,Lorenzo2018,Wauters2019,Voorden2019,Wang2019,Hatsugai2020,Lin2020,Pasquale2020,sumantopo1,Hayward2021,mondal2021,Kuno2021,Eric2022,Qidong2022,Tuloup2022}. On the experimental front, the topological Thouless pumping has been studied in the presence of disorder~\cite{Takahashi2021pump} and strongly correlated regime~\cite{walter2022breakdown}. In the attempt to extend the concept of topology to finite temperature and open quantum systems~\cite{Avron2011,Bardyn2013,Viyuela2014,Zhoushen2014,Viyuela2014_1,Nieuwenburg2014,Fleischhauer2016,Grusdt2017,Fleischhauer2018,Fleischhauer2020,Fleischhauer2021}, Thouless charge pumping has also been studied in these contexts~\cite{Fleischhauer2016,Fleischhauer2018,Fleischhauer2020}. 

{The robustness of the topological properties of a system against an open environment is important to study since quantum systems, in general, always interact with the environment. There are several studies that deal with open topological systems where they investigate the density matrix by solving the master equation~\cite{Viyuela2012,Viyuela2013,Yan2022}. We study a closed system where the topological system is a sub-system, and the rest of the system acts as an environment. The dynamics, in this case, is unitary. In our case, we consider an ensemble of uncoupled classical harmonic oscillators as the environment. In the solid state context, this realizes Einstein phonons. Our primary goal is to study the breakdown of topology in the presence of the coupling to the environment. Note that the interplay between topology and phonon has been studied before in different systems such as quantum Hall and systems with Majorana fermions~\cite{Yuval2018,Goldstein2011,Christina2018,Pavel2019}.}

\begin{figure}[b]
    \includegraphics[clip, trim={{1.7\linewidth} {1.5\linewidth} {1.7\linewidth} {1.5\linewidth}}, width=1.\linewidth]{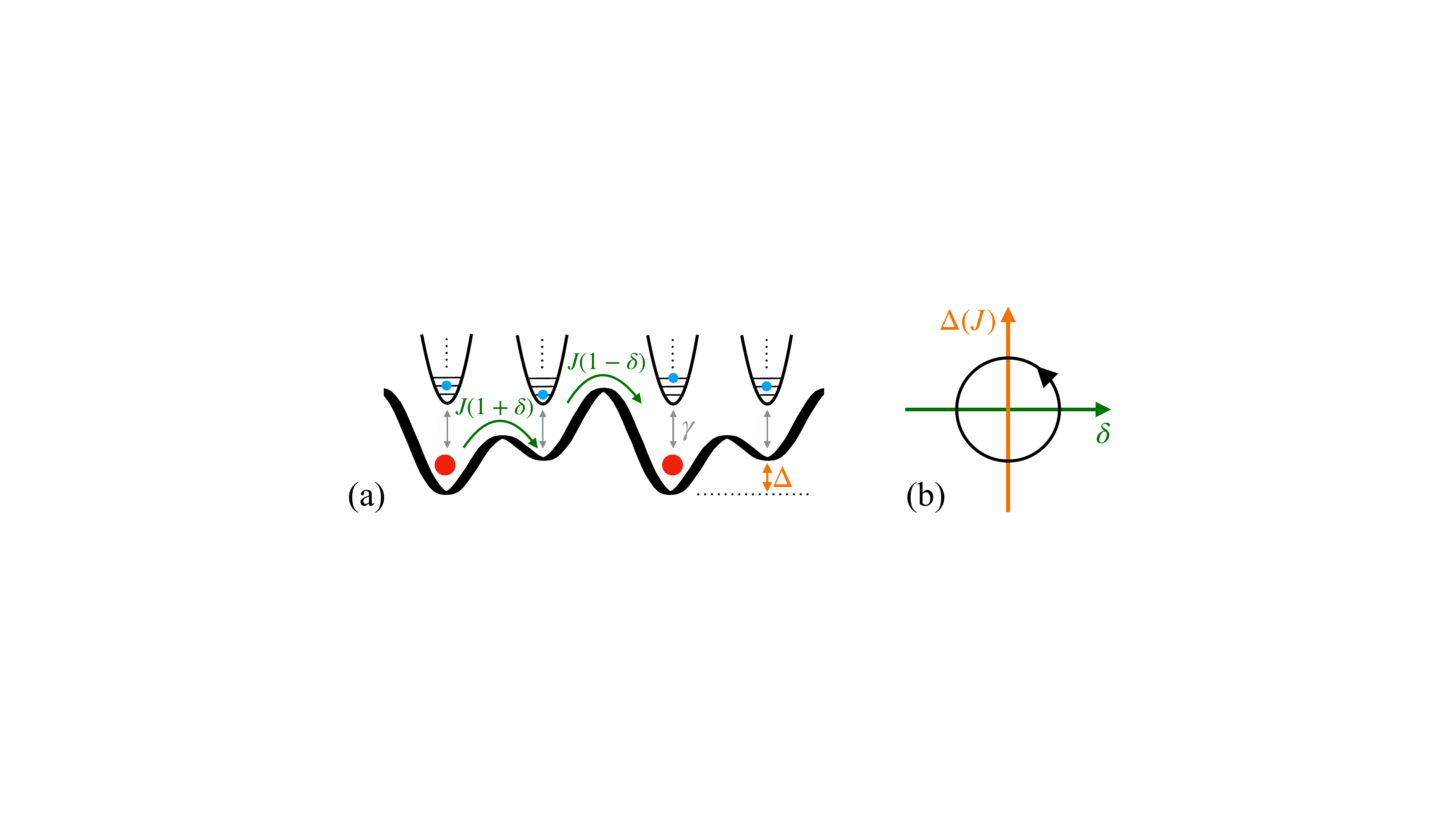}
    \caption{A pictorial representation of system parameters is shown. (a) Potential landscape at a specific point in time during a pump cycle that determines the hopping dimerization $\delta$ and staggered potential $\Delta$. The phonon bath at every site is represented by the harmonic oscillator potential, coupled to the site with strength $\gamma$. The pumping protocol is shown in (b). If the system parameter winds around the origin adiabatically, quantized pumping is expected for $\gamma=0$.}
    \label{fig:mod}
\end{figure}

One of the very crucial aspects of many-body physics is the interaction with the lattice degree of freedom. Phonons arise naturally as the lattice vibrations at any finite temperature in a real material. The electron-phonon (e-ph) coupling can cause instabilities in the metallic state leading to a plethora of important phenomena, e.g., polaron formation~\cite{Frohlich1954}, superconductivity, and charge density waves~\cite{Hirsch1983,McKenzie1996,Bursill1998,altlandbook2010,emin2012Polarons,BatrouniHolstein2019}. The phononic degrees of freedom show a drastic effect on the dynamical properties as a result of the exchange of energies between the electronic and phononic sector~\cite{Hiroaki2012, Filippis2012,Fabian2015,Hiroshi2017,Zhongkai2019,Stolpp2020,David2021,huang2021}. A great deal of research has been done to analyze the effect of phonons on dynamics of charge density wave and Mott phases~\cite{Hiroaki2012,Filippis2012,Hiroshi2017, Stolpp2020, David2021}, Bloch oscillations~\cite{Zhongkai2019,huang2021}, equilibration of excited charge carriers~\cite{Fabian2015} and thermalization~\cite{Jan2016}. The formation of and evolution of polarons have also been extensively studied~\cite{Lev2011,Guangqi2013, Werner2015, Zhongkai2017,LiChung2017,Denis2012}. One theoretical model that represents such strongly correlated e-ph systems where the electrons are coupled to the local phonons is the Holstein model~\cite{Holstein1959}.

In this work, we analyze the stability of topology where the system is coupled to a phononic environment. The transport of charge in Thouless charge pumping is studied using the Rice-Mele model~\cite{Rice1982} subjected to optical phonons, which {we dub} the Rice-Mele-Holstein model. We use the semi-classical method known as multi-trajectory Ehrenfest (MTE)~\cite{Ehrenfest,Tully1998,Kirrander2020,David2022} and time evolve the initial state with the phonons being in the ground state. The trajectory-based mixed quantum-classical dynamics was traditionally introduced to study electron-nuclear systems, which works best in the adiabatic limit of phonons with oscillator frequencies smaller than the electronic bandwidth~\cite{Ehrenfest,Tully1998,Kirrander2020,David2022}.

In our analysis with the MTE method, we see a breakdown of topological charge pumping (TCP) in the presence of phonons. At resonance, when the phonon frequency and pumping frequency are equal, the TCP breaks for any finite value of e-ph coupling. In the minimal e-ph coupling limit, the TCP eventually disappears at the higher pump cycles. It takes a finite e-ph interaction before the TCP breaks down away from the resonance. In other words, the TCP is stable at sufficiently finite e-ph coupling in this case. The breakdown of TCP is complimented with a gap-closing in the spectrum of the instantaneous effective Hamiltonian during the evolution. 

Apart from quantized pumping, there exist parameter regimes where non-quantized positive and negative pumping happen. The analysis of an effective pumping path very efficiently explains different aspects, such as the direction of pumping and breakdown of pumping. Here, the effective pumping path is the modified Rice-Mele path due to the coupling between the phonon position, which changes dynamically, and onsite electron density. The direction of winding around the origin and the origin crossing by the effective pumping path determine the direction and breakdown of pumping, respectively. We present a stability diagram for the quantized pumping as a function of e-ph coupling strength and pumping period. {For a fixed e-ph coupling (small), the diagram displays a signature of re-entrant TCP as a function of time period of pumping, i.e., the quantization of the pumped charge reappears after losing quantization near the resonance.}

{In summary, the main findings of this paper are - (i) We see a robust quantized pumping with finite electron-phonon coupling for a wide range of parameters which eventually disappear at the large electron-phonon coupling. (ii) A resonance effect exists where the quantized pumping breaks down for any finite electron-phonon coupling. (iii) The system exhibits a phenomenon of negative pumping, where the charge flows opposite to the driving, in some parameter regime yet this occurs outside the quantized regime.}

The rest of the paper is organized as follows. In Sec. \ref{sec:mod}, {we review} the model and the methods used in this work. There we introduce the Rice-Mele model that defines the Thouless pumping protocol and extend it with a Holstein-like coupling to local phonons. Next, we briefly explain the trajectory-based semi-classical MTE method, which is employed to perform the time evolution, and describe the observables used to analyze the physics. The concept of an effective pumping path is introduced at the end of this section. The discussion of results starts from Sec.~\ref{sec:res}. In this section, we explain the breakdown of TCP and non-quantized pumping in terms of the pumped charge, effective pumping paths, and instantaneous eigenvalue spectra at resonance. In Sec.~\ref{sec:outres}, we do a similar analysis as Sec.~\ref{sec:res} when the system is out of resonance. Then, we describe the stability diagram obtained as a function of the time period of pumping and e-ph coupling strength in Sec.~\ref{sec:pd}. Finally, we draw conclusions in Sec.~\ref{sec:con}.

\section{Model and Method}\label{sec:mod}
\subsection{The Rice-Mele model}
The Thouless pumping protocol can be defined by the Rice-Mele model~\cite{Rice1982}, a time-dependent Hamiltonian with time ($t$) varying superlattice potential, given by
\begin{align}\label{eq:rm}
    \hat{H}_{RM}(t) & = \sum_{i=1}^L -J(1+(-1)^i\delta(t))(\hat{c}_i^\dagger \hat{c}_{i+1} + \rm{h.c.}) \nonumber\\
                & + \sum_{i=1}^L (-1)^i\frac{\Delta(t)}{2}\hat{n}_i
\end{align}
that changes the hopping amplitude and onsite potential with time. Here $\hat{c}_i$ ($\hat{c}^\dagger_i$) and $\hat{n}_i$ are the fermionic (say electrons) annihilation (creation) and onsite number operators respectively at site $i$ in a system consisting of $L$ sites. $\delta (t)$ and $\Delta(t)$ are the hopping dimerization and onsite staggered potential, respectively, that vary with $t$ as,
\begin{align}
    \delta(t) & = A_\delta\sin{\left(\frac{2\pi t}{T} + \phi\right)} \nonumber\\ 
    \Delta(t) & = A_\Delta\cos{\left(\frac{2\pi t}{T}+ \phi\right)},
\end{align}
where $T$ is the pumping period and $\tau= t/T$ can be considered as the pumping parameter. $\phi$ is an offset of the pumping protocol which is considered to be $\phi = \pi/2$ in this work, realizing $\Delta(t)=0$ and $\delta(t)>0$ at $t=0$, known as the Su-Schrieffer-Heeger (SSH)~\cite{Su1979} limit. An example of the potential landscape is shown in Fig.~\ref{fig:mod}(a) for some $t$ which demonstrates the hopping dimerization and staggered potential. The unit cell of the lattice consists of two sublattices $A$ and $B$, implying the existence of two bands. For finite $A_\delta$ and $A_\Delta$, the band gap is always finite for all $t$ at half-filling and $t$ defines a pumping trajectory that winds around the gap-closing point at $\Delta = \delta = 0$ as shown in Fig.~\ref{fig:mod}(b). For an adiabatic change of $\tau$ (large enough $T$), the topological nature ensures the pumping of a quantized amount of charge $Q$ in a pump cycle. We can quantify the total number of charges pumped during the time evolution by integrating the current at a particular bond between two sites (the first odd bond is considered in our results) as
\begin{equation}
Q(t) = \int_0^t I_{i,i+1}(t) dt
\end{equation}
where 
\begin{equation}
I_{i,i+1}(t) = -2\Im J_{i,i+1} (t) \langle \Psi(t)| \hat{c}_{i+1}^\dagger \hat{c}_i |\Psi(t)\rangle
\end{equation}
is the current at time $t$ at the bond between site $i$ and $i+1$ which has time-dependent hopping amplitude $J_{i, i+1} = J(1+(-1)^i\delta(t))$.

\subsection{The Rice-Mele-Holstein model}
The coupling between charge carriers (electrons) and optical phonons extends the Rice-Mele model (Eq.~\eqref{eq:rm}) to the Rice-Mele-Holstein model. The Holstein-like coupling is defined by the dispersion-less phonons, also known as Einstein phonons, coupled to the local electronic state, given by
\begin{equation}
  \hat{H}(t) = \hat{H}_{RM}(t) + \hat{H}_{ph} + \hat{H}_{e-ph}.
\end{equation}
Here $\hat{H}_{ph}$ and $\hat{H}_{e-ph}$ are the phonon and e-ph coupling part of the Hamiltonian, which are given by
\begin{equation}
  \hat{H}_{ph} = \hbar \omega \sum_{i=1}^L \hat{b}_i^\dagger \hat{b}_i
\end{equation}
and 
\begin{equation}
  \hat{H}_{e-ph} = -\gamma \sum_{i=1}^L (\hat{b}_i^\dagger + \hat{b}_i) \hat{n}_i.
\end{equation}
$\hat{b}_i$($\hat{b}^\dagger_i$) are the bosonic annihilation (creation) operators of the phonons at site $i$. $\omega$ and $\gamma$ are the phonon frequency and e-ph coupling strength, respectively. The e-ph coupling is shown in Fig.~\ref{fig:mod}(a), which illustrates how the phonon bath is coupled to a site. 
 
The Thouless pump in this scenario can be studied with the real-time evolution of an initial state $|\Psi(0)\rangle$, under the influence of a time-dependent Hamiltonian $\hat{H}(t)$. In our calculation we consider $|\Psi(0)\rangle$ as 
\begin{equation}
    |\Psi(0)\rangle = |\Psi_e(0)\rangle |\Psi_{ph}(0)\rangle,
\end{equation}
where $|\Psi_e(0)\rangle$ is the ground state of $\hat{H}_{RM}(0)$ at half-filling and $|\Psi_{ph}(0)\rangle = \prod_{i=1}^{L}|\psi_{ph}^i(0)\rangle$ represents the ground state of $\hat{H}_{ph}$ where at each site $i$, phonons are in their ground state $|\psi_{ph}^i(0)\rangle$ of each individual oscillator.

\begin{figure*}
\begin{center}
\includegraphics[width=1.\linewidth]{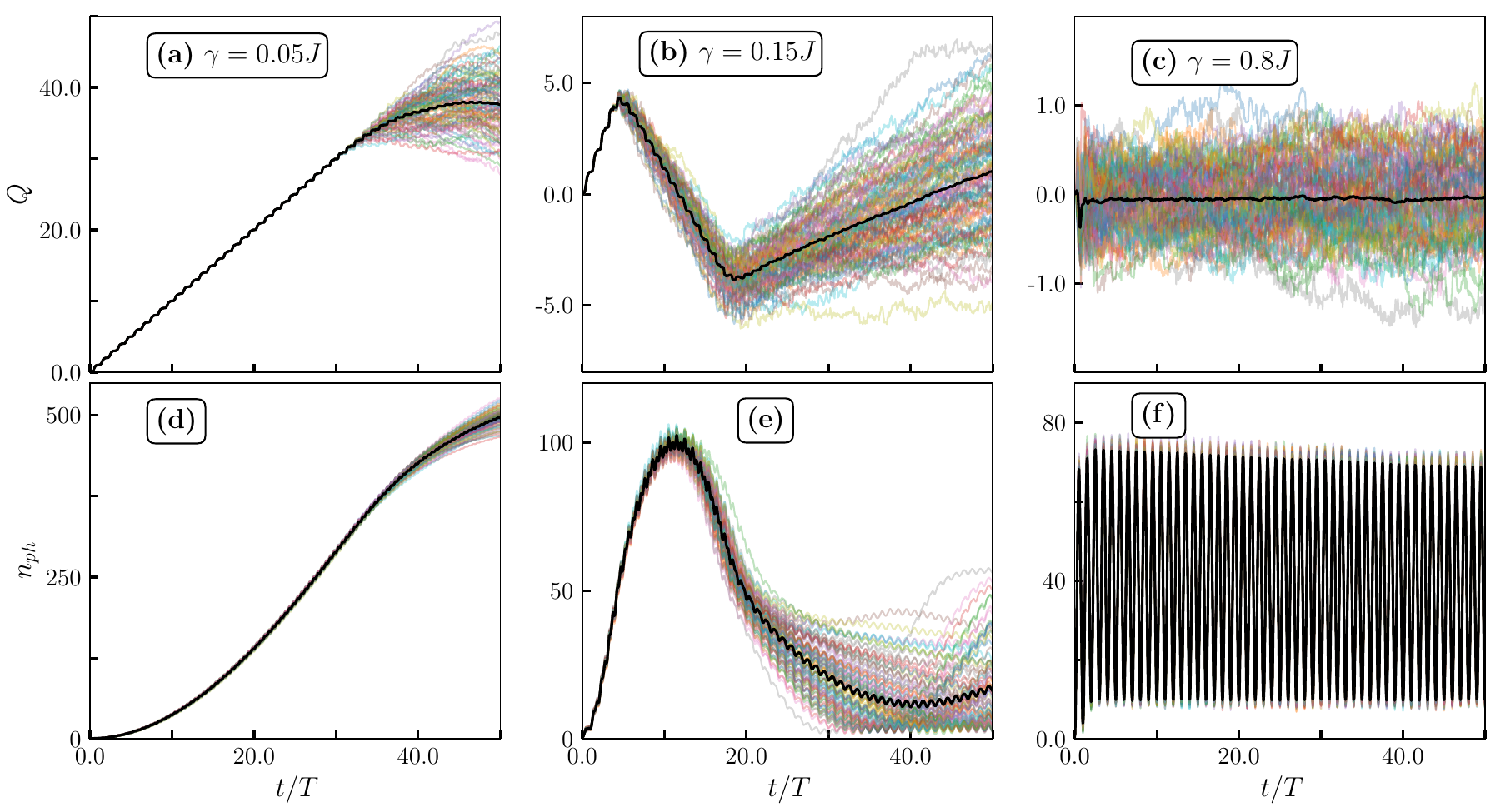}
 \end{center}
\caption{The figure shows observables obtained in the time evolution in 50 pump cycles at resonance for $N_{traj} = 100$ and $\omega = 0.1J$ with $T=2\pi/\omega$. The $Q(t)$ for all the trajectories (lines with transparent colours) and trajectory average $Q(t)$ (black line) is shown in (a-c) for $\gamma = 0.05J,~0.15J$, and $0.8J$, respectively. Similar to (a-c), we plot the phonon density $n_{ph}$ in (d-f), respectively.}
\label{fig:Q62T}
\end{figure*}

\subsection{Multi-trajectory Ehrenfest method}
In principle, the problem can be simulated using numerical methods such as exact diagonalization, Lanczos time evolution, or the time-dependent density matrix renormalization group (tDMRG) methods to capture exact dynamics for relatively smaller systems~\cite{Brockt2015,Stolpp2020,Manmana2005,Vidal2004,Schollwock2011}, yet not straightforwardly in the adiabatic regime. In this work, we employ a semi-classical approximate approach called multi-trajectory Ehrenfest (MTE) method, which treats phonons classically and averages over independent trajectories. Since we are primarily interested in the regime of small phonon frequencies where we expect the most immediate effect on the pump, we choose MTE, which is efficient and reliable in this regime~\cite{David2022}. This method simplifies our problem to such an extent that it can be solved using the single-particle eigenstates of the Hamiltonian. The classical trajectory-based MTE method is described extensively in Ref.~\cite{David2022} and in the following, we describe it briefly. To apply the MTE method the phononic operators are represented in real space via $\hat{b}_i^\dagger = \sqrt{\frac{m\omega}{2\hbar}}(\hat{x}_i + \frac{\hat{p}_i}{m\omega})$ and by using the natural length scale for the harmonic oscillators $l_0 = \sqrt{\frac{\hbar}{m\omega}} = 1$ and $\hbar = 1$, we get
\begin{equation}
\hat{H}_{ph} = \frac{\omega}{2}\sum_{i=1}^L (\hat{x}_i^2 + \hat{p}_i^2) ,
\end{equation}
ignoring the constant term $-\omega L/2$, and 
\begin{equation}
\hat{H}_{e-ph} = -\sqrt 2 \gamma \sum_{i=1}^L \hat{x}_i \hat{n}_i.
\end{equation}
Under the MTE approximation, $|\Psi(0)\rangle$ can now be evolved by initializing the different sets of phonon coordinates $\{x_i(0),p_i(0)\}$, which define different independent trajectories. The $\{x_i(0),p_i(0)\}$ are randomly drawn from the distribution given by the Wigner function $W_0$ centered around $\langle x \rangle = 0$ and $\langle p \rangle = 0$.  $W_0$ is the phase space representation of the harmonic oscillator ground state $|\psi_{ph}(0)\rangle$,  
\begin{equation}
W_0 = \frac{1}{\pi}e^{-(x - \langle x \rangle)^2 - (p - \langle p \rangle)^2}. 
\end{equation}
The electronic wave function $|\Psi_e(0)\rangle$ evolves under the Hamiltonian 
\begin{equation}\label{eq:Hel}
\hat{H}_{el} = \hat{H}_{RM}(t) - \sqrt 2 \gamma \sum_{i=1}^L x_i(t) \hat{n}_i
\end{equation}
which only depends on the dynamical position $x_i(t)$ of the phonons, $i\frac{\partial}{\partial t} |\Psi_e\rangle = \hat{H}_{el} |\Psi_e\rangle$. The phonon coordinates simultaneously propagate in phase space via the Newtonian mechanics under the influence of the clasical Hamiltonian 
\begin{equation}
H_{cl} = \frac{\omega}{2}\sum_{i=1}^L ({x}_i^2 + {p}_i^2) - \sqrt 2 \gamma \sum_{i=1}^L {x}_i \langle\Psi_e(t)|\hat{n}_i|\Psi_e(t)\rangle
\end{equation}
as, $\dot{x_i} = \frac{\partial H_{cl}}{\partial p_i},~~~ \dot{p_i} = -\frac{\partial H_{cl}}{\partial x_i}$.

The expectation value of any observable $\hat{O}$ can be calculated by averaging over all the trajectories as
\begin{equation}
    \langle \hat{O}(t) \rangle = \frac{1}{N_{traj}}\sum_{i=1}^{N_{traj}} \langle \Psi_i(t)| \hat{O} | \Psi_i(t) \rangle,
\end{equation}
where $N_{traj}$ is the number of trajectories and $|\Psi_i(t)\rangle$ is the time-evolved initial electronic wave function $|\Psi_e(0)\rangle$ at time $t$ for the $i$-th trajectory. In the following section, we present the results where we always consider $J=1$, making all the parameters unitless {and express everything in units of $J$} and $A_\Delta = 4A_\delta = 3$, defining a closed path around the gap-closing point at the origin in Fig.~\ref{fig:mod}(b). {Considering the limitations of the multi-trajectory Ehrenfest method as mentioned above, we take $\omega = 0.1J$, which is small compared to $J$.} In our numerical simulation, we always start the pumping protocol from a short negative time ($\frac{-T}{16J}$) without any phonon coupling and drive it slowly (compared to the actual pumping speed) to $t=0$ where we quench $\gamma$. The slow drive before $t=0$ leads to a smoother pumping during the execution of the pump cycles~\cite{Lorenzo2018}.

\subsection{Effective pumping path}\label{sec:eff_pump}
As mentioned above, the electronic dynamics is determined by the $\hat{H}_{el}$ given in Eq.~\eqref{eq:Hel}, which has two parts. The first part contains the control parameters ($\delta(t)$ and $\Delta(t)$) which defines a pumping path as shown in Fig.~\ref{fig:mod}(b). We expect the pumping to be positively or negatively quantized depending on the direction of the winding around the origin. The second part of the $\hat{H}_{el}$ gives rise to another source to the onsite potential on top of the staggered $\Delta(t)$ by coupling $x_i(t)$ with the onsite $\hat{n}_i$. This additional source of the onsite potential may alter the effective staggered potential between the sub-lattices leading to a deviation in the pumping path from the original one defined by $\hat{H}_{RM}$. We can quantify the effective pumping path in the ($\bar{\delta}, \bar{\Delta}$) plane where $\bar{\delta}(t) = 2\delta(t)$ is the hopping dimerization, which is unchanged and $\bar{\Delta}(t)$ is the trajectory and unit-cell averaged potential difference between two sub-lattices, given by
\begin{align}\label{eq:rm}
   & \bar{\Delta}(t) = \Delta(t) \nonumber\\
                & - \frac{2\sqrt{2}\gamma}{LN_{traj}}\sum_{i=1}^{N_{traj}}\left(\sum_{j \in \rm{even}}^L x_{i,j}(t) -  \sum_{j \in \rm{odd}}^L x_{i,j}(t)\right).
\end{align}
The following sections explain different phenomena, such as the breakdown of pumping and negative charge pumping using the effective pumping path.

N.B.: The numerical data, plotted in the figures below, are partially available in arXiv ancillary files.

\section{Resonance Condition}\label{sec:res}
While the RM model shows robust TCP, the nonzero coupling ($\gamma > 0$) with the phonons gives rise to very rich physics. We begin the analysis with the condition when the phonon frequency matches the pumping frequency ($\omega = 2\pi/T$). The $\omega$ is considered as $0.1J$ which fixes the pumping period as $T= 2\pi/\omega$. In this scenario, the dynamics are adiabatic enough in the RM limit ($\gamma = 0$) with robust quantized pumping. The question is whether this TCP survives at finite $\gamma$. To answer this, we calculate different quantities during fifty pump cycles with different e-ph coupling strengths $\gamma$. In Figs.~\ref{fig:Q62T}(a-c) we plot the $Q(t)$ for $\gamma = 0.05J,~0.15J$ and $0.8J$, respectively, for $100$ trajectories (transparent lines) where the average $Q(t)$ over the trajectories is represented by the opaque black line. We can see that even for a very small value of $\gamma=0.05J$, the TCP breaks down in the later pump cycles. For $\gamma = 0.15J$, we see a complex behavior in pumping; it changes the direction of pumping after a few cycles. For a large value of $\gamma = 0.8J$, the pumping ceases to occur. We can notice that when the quantization of pumping breaks down, the $Q(t)$ becomes trajectory dependent which fluctuates around the average value. This behavior is similar to pumping in a disorder potential~\cite{Hayward2021}, which is not surprising since each trajectory is initialized with random initial values of $\{x_i(0),p_i(0)\}$.

Interestingly, the phonon excitation is very distinct in different parameter regimes of $\gamma$. Figures~\ref{fig:Q62T}(d-f) shows the behaviors of the phonon density given by
\begin{equation}
n_{ph} = \frac{1}{L} \sum_{i=1} \left(\frac{x_i^2}{2} + \frac{p_i^2}{2}\right)
\end{equation}
for the same parameters considered in Figs.~\ref{fig:Q62T}(a-c), respectively. Even though $Q(t)$ is quantized in the first few pump cycles for smaller $\gamma$ (Figs.~\ref{fig:Q62T}(a), and \ref{fig:Q62T}(b)) $n_{ph}$ continuously increases and eventually, the quantization of $Q(t)$ breaks down. When the pumping is completely suppressed (Fig.~\ref{fig:Q62T}(c)), $n_{ph}$ oscillates around a finite number with the frequency $\omega$.

\begin{figure}
\begin{center}
\includegraphics[width=1.\linewidth]{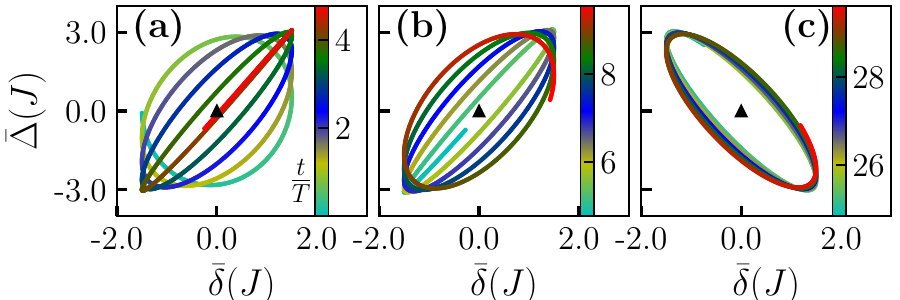}
\includegraphics[width=1.\linewidth]{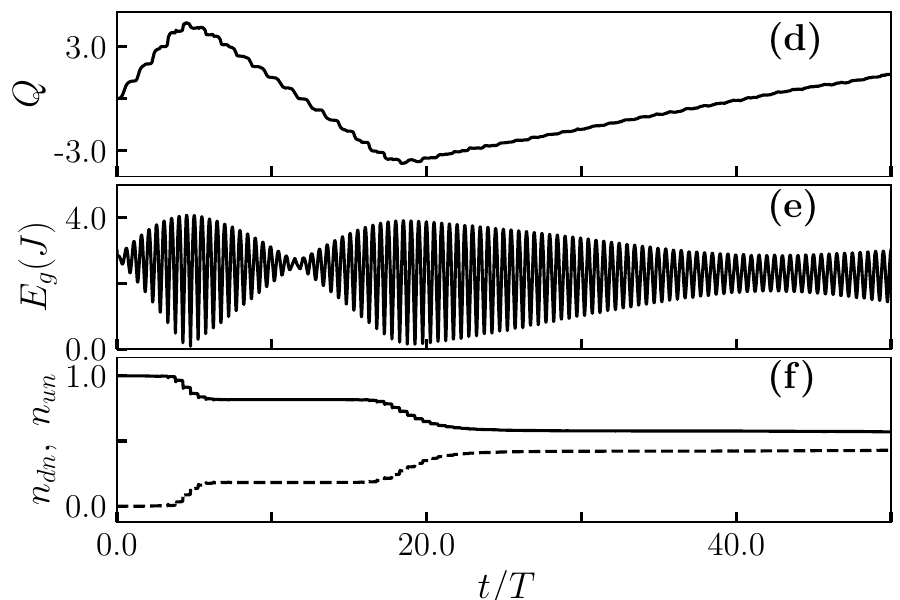}
\end{center}
\caption{Trajectory averaged quantities are displayed here at resonance for $\gamma=0.15J$, $\omega =0.1J$ and $T = 2\pi/\omega$ with $N_{traj} = 100$. The trajectory averaged effective pumping path is shown for three different time windows in (a), (b), and (c) corresponding to Fig.~\ref{fig:Q62T}(b), which is also shown in (d) for reference. The black triangles mark the origins where the gap should vanish. The direction of winding around the origin signifies the direction of pumping. The $E_g(t)$ is shown in (e) and $n_{\rm{dn}}(t)$ (solid line) and $n_{\rm{up}}(t)$ (dashed line) are plotted in (f).}
\label{fig:62Tenr}
\end{figure}

\begin{figure}
\begin{center}
\includegraphics[width=1.\linewidth]{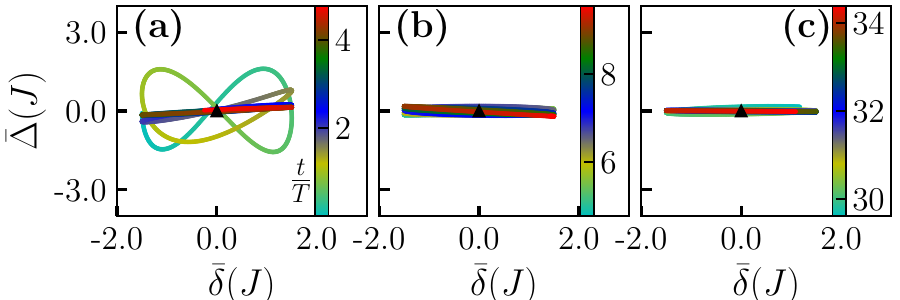}
\includegraphics[width=1.\linewidth]{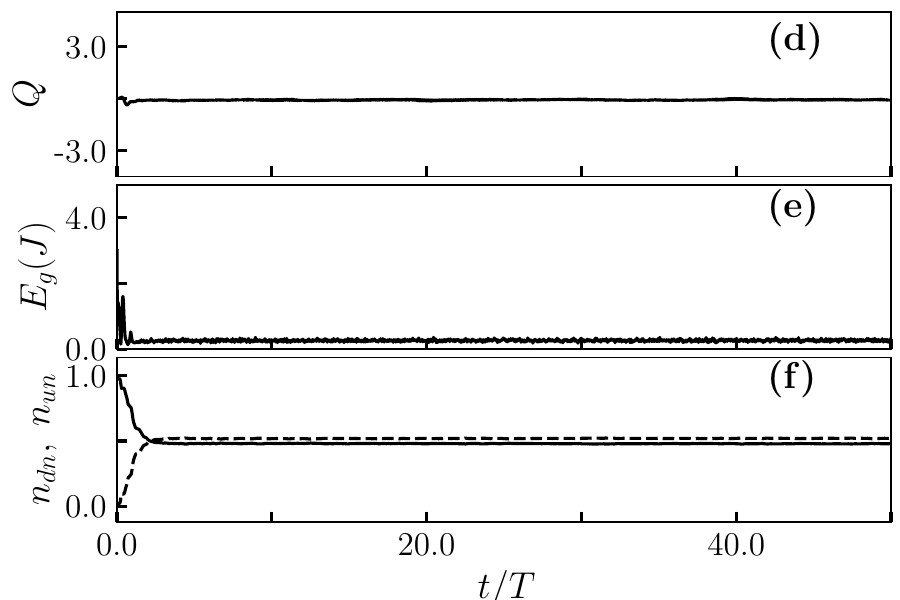}
\end{center}
\caption{The figure explains breakdown of pumping at resonance for $\gamma=0.8J$, $\omega =0.1J$ and $T = 2\pi/\omega$. Different quantities are plotted after averaging over $100$ trajectories. The effective pumping path is shown for three different time windows in (a), (b), and (c) corresponding to Fig.~\ref{fig:Q62T}(c), which is also shown in (d) for reference. The black triangles mark the origins where the gap should vanish. The frequent crossing of origin by the effective pumping path signifies the breakdown of TCP. The $E_g(t)$ is shown in (e) and $n_{\rm{dn}}(t)$ (solid line) and $n_{\rm{up}}(t)$ (dashed line) are plotted in (f).}
\label{fig:62Tenr1}
\end{figure}

\begin{figure*}
\begin{center}
\includegraphics[width=1.\linewidth]{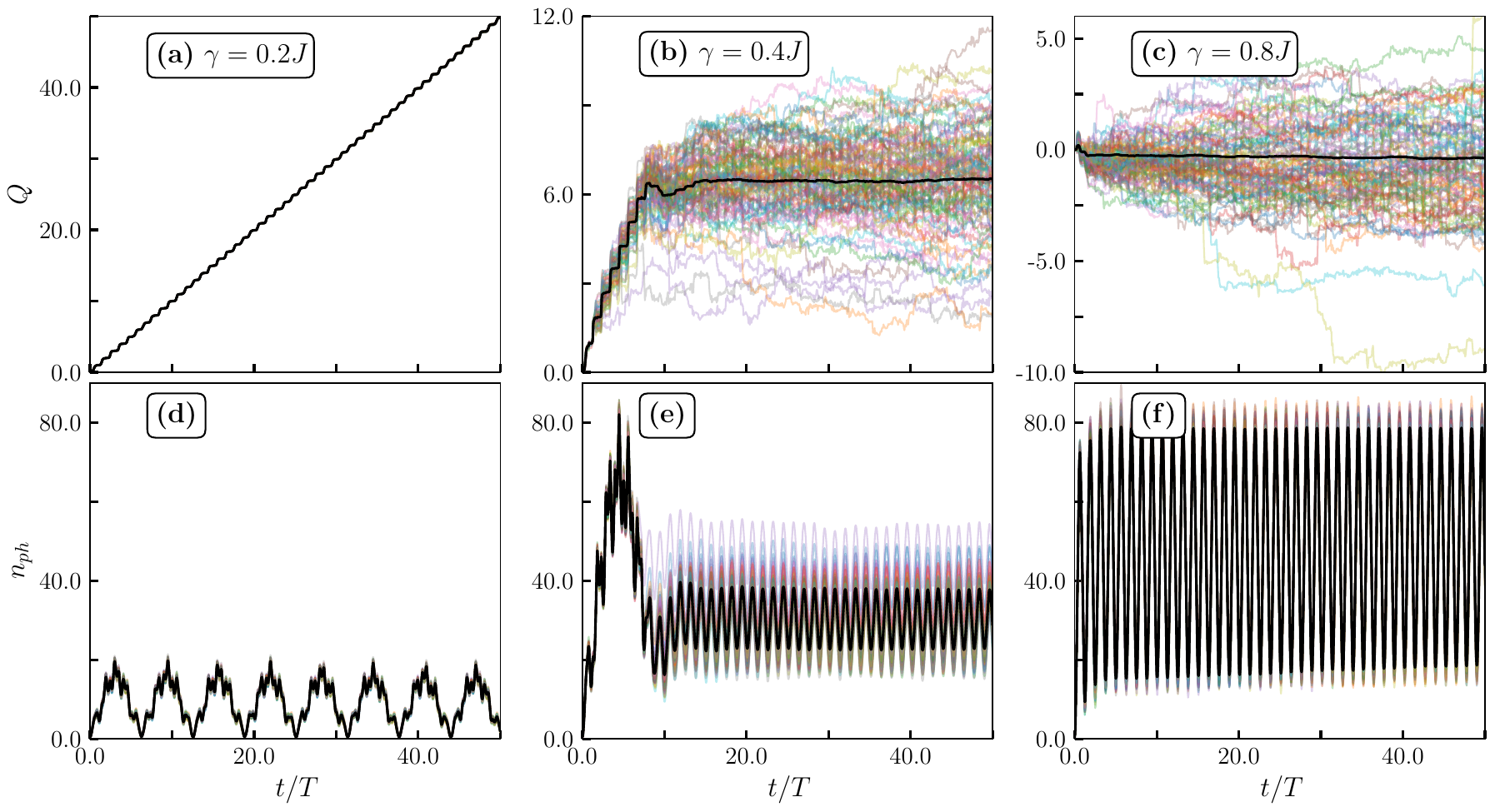}
 \end{center}
\caption{The figure shows observables obtained in the time evolution in 50 pump cycles away from resonance for $N_{traj} = 100$ and $\omega = 0.1J$ with $T=50/J$. The $Q(t)$ for all the trajectories (lines with transparent colours) and trajectory average $Q(t)$ (black line) is shown in (a-c) for $\gamma = 0.2J,~0.4J$, and $0.8J$, respectively. Similar to (a-c), we plot the phonon density $n_{ph}$ in (d-f), respectively.}
\label{fig:Q50T}
\end{figure*}

\begin{figure}
\begin{center}
\includegraphics[width=1.\linewidth]{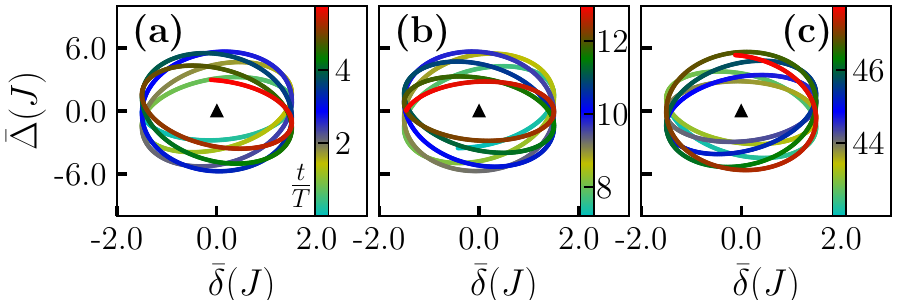}
\includegraphics[width=1.\linewidth]{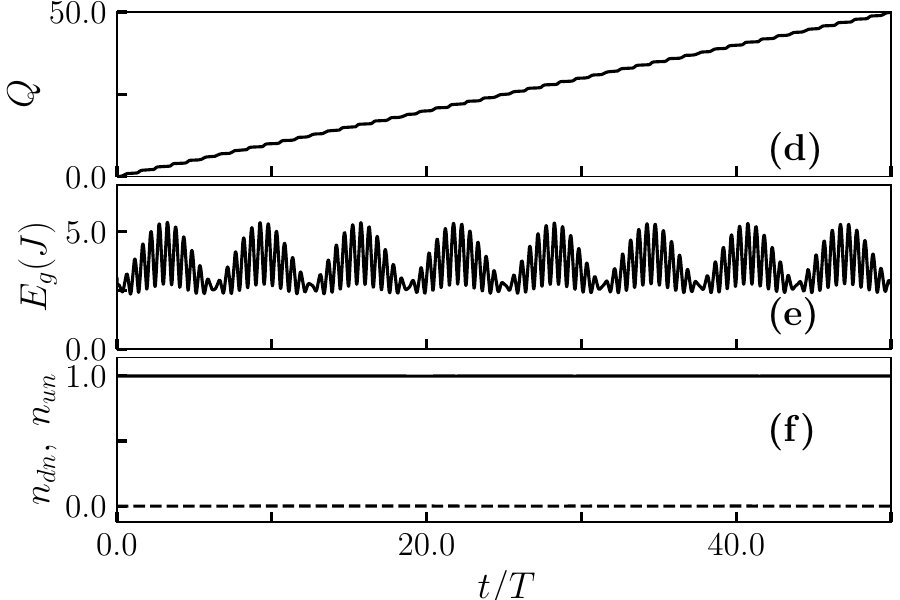}
\end{center}
\caption{Different quantities are displayed here after averaging over trajectories with $N_{traj} = 100$ to analyze the quantized pumping at finite $\gamma=0.2J$ away from resonance where $\omega =0.1J$ and $T = 50/J$. The effective pumping path is shown for three different time windows in (a), (b), and (c), corresponding to Fig.~\ref{fig:Q50T}(a), which is also shown in (d) for reference. The black triangles mark the origins where the $E_{g}$ should vanish. The $E_g(t)$ is shown in (e) and $n_{\rm{dn}}(t)$ (solid line) and $n_{\rm{up}}(t)$ (dashed line) are plotted in (f).}
\label{fig:62Tenr2}
\end{figure}
\begin{figure}
\begin{center}
\includegraphics[width=1.\linewidth]{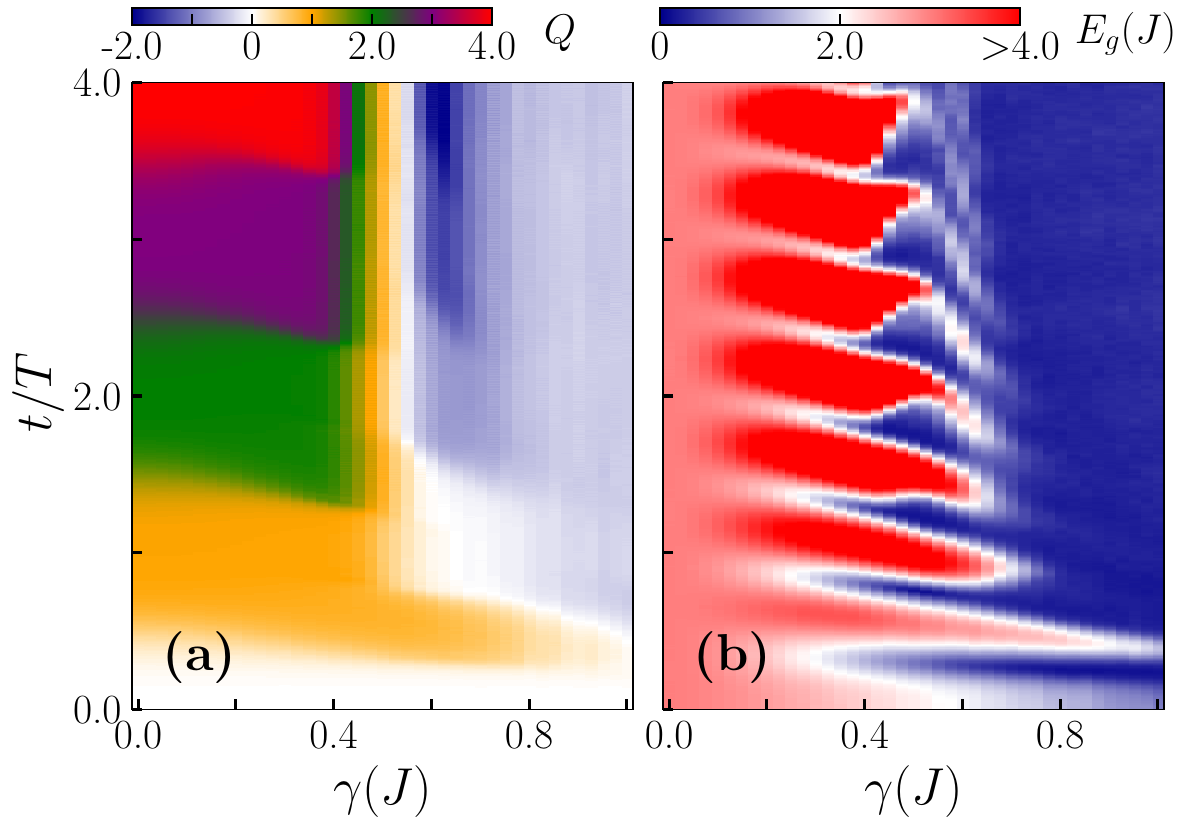}
 \end{center}
\caption{(a) The trajectory averaged $Q(t)$ and (b) $E_{gap}(t)$ are shown away from resonance with $T=50/J$, $\omega = 0.1J$ and $N_{traj} = 200$. The figure portrays the breakdown of quantized pumping near $\gamma \sim 0.4J$.}
\label{fig:gap}
\end{figure}

\begin{figure}
\begin{center}
\includegraphics[width=1.\linewidth]{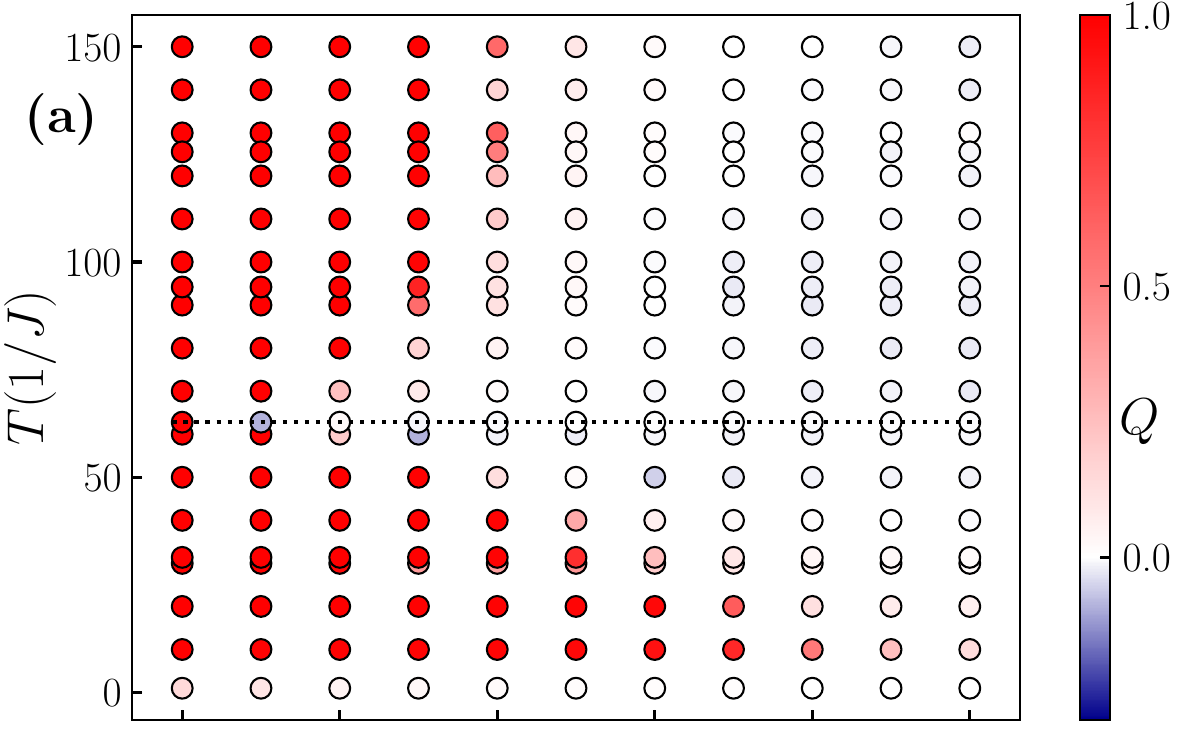}
\includegraphics[width=1.\linewidth]{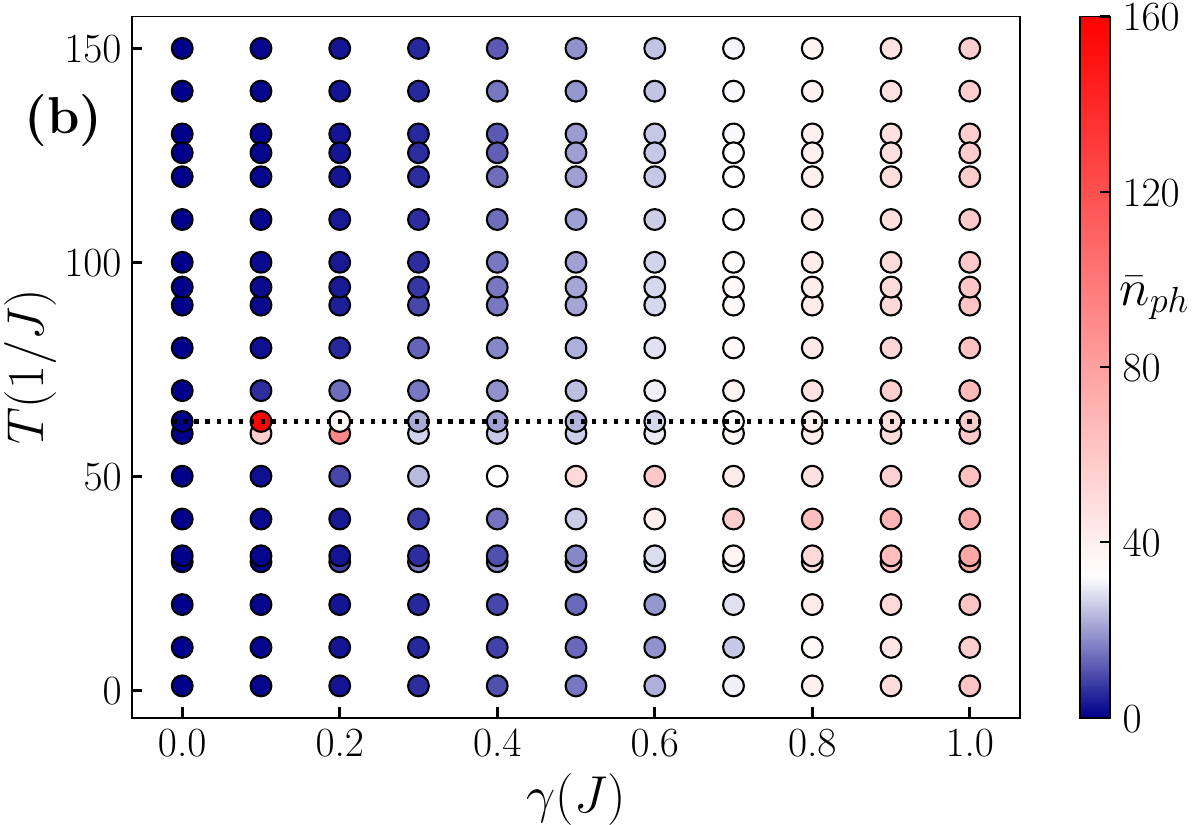}
 \end{center}
\caption{The figure represent a stability diagram in $T$ vs $\gamma$ plane depending on the trajectory averaged (a) $Q$ in a cycle and (b) $\bar{n}_{ph}$. Here we consider $N_{traj} = 100$ and average the $Q$ and $\bar{n}_{ph}$ over fifty pump cycles. The dashed lines represent the time period of phonons ($\frac{2\pi}{\omega}$).}
\label{fig:pd}
\end{figure}

\subsection{Negative charge pumping}
The intricate evolution of $Q(t)$ seen in Fig.~\ref{fig:Q62T}(b) is observed in a finite range of $\gamma$. To explain the multiple changes in pump direction, we analyze the effective pumping path. As discussed in Sec.~\ref{sec:eff_pump}, the instantaneous value of $x_i(t)$ can modify the effective pumping path. The pumping path defined by ($\bar{\delta}(t), \bar{\Delta}(t)$) is plotted for different time segments of the plot shown in Fig.~\ref{fig:Q62T}(b), where the pumped charge is first positive (quantized), then negative (non-quantized), and becomes positive (non-quantized) again with time, in Figs.~\ref{fig:62Tenr}(a), (b), and (c), respectively. The colour-bar represent the number of pump cycle ($t/T$). We can see from Figs.~\ref{fig:62Tenr}(a), (b), and (c) that the path winds the origin (solid black triangle) first in anti-clockwise, then clockwise and then anti-clockwise direction again, respectively, which explains the direction of pumping. To check further why the quantization breaks down, we look into the trajectory averaged gap in the spectrum of instantaneous $\hat{H}_{el}(t)$ given by
\begin{equation}
E_g(t) = E_{L/2+1}^{el}(t) - E_{L/2}^{el}(t),
\end{equation}
where $\{E_{\alpha}^{el}(t)\}$ is the time-dependent energy spectrum of $\hat H_{el}(t)$ along with the occupancy of the lower ($n_{\rm{dn}}$) and upper ($n_{\rm{up}}$) band of it with time. We can already see from Fig.~\ref{fig:62Tenr}(a) that, at the end of the first few cycles, the effective pumping path crosses the origin, where the gap in the spectrum of $\hat H_{el}(t)$ should close. The $E_g(t)$ plotted in Fig.~\ref{fig:62Tenr}(d) exactly captures this gap-closing. Until this point, the $n_{\rm{dn}}$ (solid line) and $n_{\rm{up}}$ (dashed line) shows that particles in $|\Psi(t)\rangle$ completely occupy the lower band giving rise to the quantized pumping in this time segment. However, at $E_g=0$, partial excitation to the upper band happens non-adiabatically, which can be inferred from the finite $n_{\rm{up}}$. After this first gap-closing point, the $E_g(t)$ becomes finite again and since the effective pumping path winds in the opposite direction in this segment (Fig.~\ref{fig:62Tenr}(b)), it results in the negative charge pumping. The partial occupancy of the lower band breaks the quantization. A similar gap closing is detected again during the evolution, which populates the upper band further. The anti-clockwise winding of the effective pumping path (Fig.~\ref{fig:62Tenr}(c)) suggests a positive pumping (non-quantized) since the $E_g(t)$ is finite in the later times.

\subsection{Breakdown of pumping}
To compare the previous situation with the case shown in Fig.~\ref{fig:Q62T}(c) where the pumping is entirely absent, we do a similar analysis in Fig.~\ref{fig:62Tenr1}. Here, the phonon coupling drastically perturbs the staggering nature of the potential, and the effective pumping path becomes flat along $\bar{\delta}$ axis which crosses the origin in every pump cycle (Figs.~\ref{fig:62Tenr1}(a-c)). As a result, the $Q(t)$ does not change with time. The vanishing $E_g(t)$, shown in Fig.~\ref{fig:62Tenr1}(e), signifies a non-adiabatic dynamics under $\hat H_{el}(t)$ leading to breakdown of pumping. In this region, the $E_g(t)$ is not exactly zero, which is a finite size effect, and expected to be zero in the thermodynamic limit. The $n_{\rm{dn}} \sim n_{\rm{up}}$ shown in Fig.~\ref{fig:62Tenr1}(f) also suggest no change in $Q(t)$ as the $|\Psi(t)\rangle$ is equally mixed in both the bands.

\section{Out of resonance Condition}\label{sec:outres}
Moving away from resonance, we encounter different physical properties. In this regime, the pumping is found to be quantized at a sufficiently finite value of $\gamma$. Ultimately, with increasing $\gamma$, the TCP breaks down. We analyze the same quantities studied in the previous section to capture the physics. In Figs.~\ref{fig:Q50T}(a-c), we plot $Q(t)$ for $\gamma = 0.2J,~0.4J$, and $0.8J$, respectively, for $100$ trajectories (transparent lines) where the average $Q(t)$ over the trajectories is represented by the opaque black line. Even though the initial $\{q_i, p_i\}$ are different for different trajectories, for finite $\gamma = 0.2J$, we see quantized $Q(t)$ all the way to the 50th pump cycle and $Q(t)$s for all the trajectories merge together which implies a robust TCP at finite phonon coupling $\gamma$. 

We observe different effects in the dynamics by increasing $\gamma$ to $0.4J$. In this case, the charge pumping breaks down after a few pump cycles, and $Q(t)$ is not quantized anymore in the few initial pump cycles that still exhibit finite pumping. For different trajectories, $Q(t)$ shows a substantial fluctuation from the average value as a result of topologically unprotected dynamics, which stems from the breakdown of TCP and largely depends on the initial conditions. Further increasing $\gamma = 0.8J$, the pumping breaks down from the beginning, and $Q(t)$ for all the trajectories show a large fluctuation similar to the $\gamma=0.4J$ case.

\subsection{Quantized pumping with phonons}
Unlike the resonance case where the TCP breaks down for small $\gamma$ (Fig.~\ref{fig:Q50T}(a)) at later pump cycles, in this case, TCP may remain stable at small $\gamma$. We confirm the stability by analyzing the behavior of the $n_{ph}(t)$. Figures~\ref{fig:Q50T}(d-f) shows the $n_{ph}(t)$ for the same parameters considered in Figs.~\ref{fig:Q50T}(a-c), respectively. We see that $n_{ph}(t)$ periodically reaches zero after every few pump cycles (Fig.~\ref{fig:Q50T}(e)) for the $\gamma = 0.2J$ case. Also, the extreme value of $n_{ph}(t)$ is comparatively smaller (compare with Fig.~\ref{fig:Q62T}(d)). This behavior is completely different from the resonance case where $n_{ph}$ keeps increasing in the initial pump cycles where pumping is quantized (compare with Figs.~\ref{fig:Q62T}(d) and (e)). This regular oscillatory behavior of the $n_{ph}$ guarantees the same dynamics for further pump cycles beyond the $50$th cycle, and TCP survives. In Fig.~\ref{fig:Q50T}(f), for the first few cycles where pumping is non-zero, the trend of $n_{ph}(t)$ looks similar to the $\gamma=0.2J$ case. However, the features look comparable to the resonance case when the pumping is absent (compare Fig.~\ref{fig:Q50T}(f) and Fig.~\ref{fig:Q62T}(f)). 

The effective pumping path for $\gamma=0.2J$ also suggests a robust winding which is shown in Figs.~\ref{fig:62Tenr2}(a-c). We can see that the pumping path evolves similarly for different time windows and does not have any trend to collapse towards the origin, indicating robust TCP. The $Q(t)$, $E_g(t)$, $n_{\rm{dn}}$ and $n_{\rm{up}}$ are plotted in Figs.~\ref{fig:62Tenr2}(d-f), respectively for this case. The finite $E_g(t)$ and $n_{\rm{dn}} =1$ ($n_{\rm{up}} =0$) at all time illustrate the adiabatic dynamics under $\hat{H}_{el}(t)$ securing the quantization of pumping.

To summarize, we observe a significant difference at small $\gamma$ comparing pumping at resonance and away from resonance. The behavior of $n_{ph}(t)$ and effective pumping path suggests stable pumping over the simulated time window. We stress that this does not rule out a change in behavior at very long times beyond the reach of simulations.

\subsection{Breakdown of pumping}
As mentioned above, for larger $\gamma$ the TCP breaks down. To analyze the breakdown of pumping with increasing $\gamma$, we again pay attention to $E_g(t)$. We show the trajectory averaged $Q(t)$ and $E_g(t)$ in Fig.~\ref{fig:gap}(a) and (b), respectively, for the first four pump cycles with $T=50/J$. Here, we consider $N_{traj} = 200$ to calculate the average. Comparing both figures, we can see that $E_g(t)$ starts to vanish at certain times near the critical region ($\gamma\approx 0.4J$) and finally vanishes during the entire pump cycle for larger $\gamma$. The vanishing of $E_g$ with time makes the dynamics under $\hat{H}_{el}$ non-adiabatic, and as a result, the TCP breaks down. Note that the negative charge pumping is also present in the off-resonant case (see the deep blue region of Fig.~\ref{fig:gap}(a)).

\section{Stability Diagram}\label{sec:pd}
Until now, we have considered only two time periods of pumping ($T=\frac{2\pi}{0.1J}$ and $50/J$) in our analysis to detect the breakdown of TCP. Now we uncover this phenomenon by varying $T$. To this end, we perform the calculation for different $\gamma$ and $T$ for the first fifty pump cycles with $N_{traj} = 100$ and calculate the trajectory averaged $Q$ in a pump cycle by averaging the pumped charge in all the cycles. We portray the result in Fig.~\ref{fig:pd}(a) as a function of $\gamma$ and $T$, where the colour bar represents the average $Q$ in a cycle. The properties of the stability diagram are discussed below.

In the limit of very small $T\sim 1/J$, the TCP is always absent due to a very fast pumping speed which leads to a non-adiabatic evolution. For higher values of $T$, the TCP breaks down after a critical $\gamma$ except at resonance (marked by the dashed line), where it breaks down immediately for $\gamma>0$. An important feature of the stability diagram is the re-entrance of quantized pumping as a function of $T$ in a certain parameter regime. For example, with $\gamma = 0.2J$, the quantization of pumping occurs in the order of non-quantized-quantized-non-quantized-quantized manner as $T$ increases. The re-entrance of TCP with increasing $T$ is clearly a consequence of the resonance at  $T=2\pi/\omega$. Simulations with other values of small $\omega$ exhibit qualitatively similar physical phenomena at the resonance and away from the resonance (not shown).

We also calculate the time averaged phonon density given by
\begin{equation}
    \bar{n}_{ph} = \frac{1}{t}\int_0^t n_{ph} dt
\end{equation}
and shown in Fig.~\ref{fig:pd}(b) as a function of $\gamma$ and $T$. Similar to the previous calculation, $\bar{n}_{ph}$ is averaged over $N_{traj}=100$ trajectories and fifty pump cycles. The result complements the stability diagram of $Q$ (Fig.~\ref{fig:pd}(a)) very well. We can notice that the phonon excitation in the TCP region is small compared to the TCP-broken region. Around the resonance ($T =2\pi/\omega$), marked by the dashed line, the $\bar{n}_{ph}$ grows faster for smaller $\gamma$, which is responsible for the early breakdown in this region.

\section{Conclusions}\label{sec:con}
In conclusion, we have studied the Thouless charge pumping in the presence of optical phonons. The Rice-Mele pumping protocol extended with Holstein-like coupling to the local dispersionless phonons is used. We considered the initial conditions where the sub-systems are decoupled, and the phonons are at their ground state.  We have utilized a semi-classical approach known as the multi-trajectory Ehrenfest method to analyze the system's dynamics where the electrons are treated quantum mechanically, and phonon trajectories are evolved classically.

The analysis reveals a breakdown of quantized pumping induced by the phonons for any finite value of the e-ph coupling when the phonon frequency and pumping frequency match. Moreover, in this case, for smaller e-ph coupling, a non-quantized positive and negative pumping is observed. The direction of pumping is accurately explained using the effective pumping path, which is the modified Rice-Mele path due to the coupling of phonon position with the electronic density. The effective pumping path is found to be winding around the origin in anti-clockwise and clockwise directions for the time windows with positive and negative pumping, respectively. The non-adiabatic nature of the dynamics affects the quantization of pumping in this regime. The non-adiabaticity is visible from the gap in the energy spectrum of the instantaneous Hamiltonian. When the pumping vanishes at higher values of e-ph coupling, the effective pumping path is seen to cross the origin in every cycle during the evolution, which explains the breakdown.

When the pumping period is out of resonance, the phonon-induced breakdown of the quantized pumping still exists. Yet, a parameter regime of robust quantized pumping exists before it breaks down at the larger e-ph coupling. The periodically oscillating phonon number evolution and non-shrinking stable effective pumping path signal the robustness of quantization in this case. The adiabatic nature of the pumping does not hold after the breakdown of pumping for larger e-ph coupling, which is visible from the energy spectrum of the instantaneous Hamiltonian.

We have obtained a stability diagram as a function of e-ph coupling and the time period of pumping. A wide region of quantized pumping is observed in the parameter space of finite e-ph coupling. As a result of resonance, a re-entrant behavior of the quantized pumping is observed where at a fixed e-ph coupling, the quantization reappears after a breakdown near the resonance as a function of the time period of pumping.

These results lead to an avenue of further analysis which worth studying. One can do a similar analysis for quantum phonons. The system can be simulated using the tDMRG method with local basis optimization technique~\cite{Brockt2015,Stolpp2020}. Starting the pump from different initial states may lead to interesting outcomes. One should further consider scenarios of stable quantized pumping even in the presence of phonons. A half-filled Holstein model, for instance, can host a charge-density wave state~\cite{Hirsch1983,McKenzie1996,Bursill1998}, depending on parameters, with a many-body gap. One can also extend the model for the phonons with dispersion. {The addition of dispersion allows the phonons to transport energy to other sites, whereas local oscillators can only do that via electrons. The sign of the dispersion may also matter as the minimum changes its position in $k$-space from $k=0$ to $k=\pi$ with the sign.}

\section{Acknowledgement}
{We thank M. ten Brink for the helpful discussions. This research was funded by the Deutsche Forschungsgemeinschaft (DFG, German Research Foundation) via Research Unit FOR 2414 under Project No. 277974659.}

\bibliography{references}
\clearpage



\begin{titlepage}
\centering
{\Large\bfseries Erratum:}

{\large\bfseries Phonon-induced breakdown of Thouless pumping in the Rice-Mele-Holstein model}
\vspace{1cm}
\end{titlepage}

Here we rectify a mistake in the implementation of the semi-classical approach of our article. The correction leads to a quantitative change in the results. However, qualitatively the physics extracted from the numerical data remains valid. The interpretation, discussion and conclusions of the article are not affected by the error. In the following, we clarify the mistake in detail. Then we update the results with rectified numerical data and mention the changes that should be done to the manuscript.


In order to evolve the phonon coordinates $\{x_i(t), p_i(t)\}$ simultaneously with the evolution of the quantum mechanical electronic part, we can use the exact solution of the equations of motion
\begin{equation}
    \dot{x_i} = \frac{\partial H_{cl}}{\partial p_i},~~~ \dot{p_i} = -\frac{\partial H_{cl}}{\partial x_i}
    \label{eq:eqm}
\end{equation}
as,
\begin{align}
 x_i(t + \delta t) = &  x_i(t) {\rm{cos}}(\omega \delta t) + p_i(t){\rm{sin}}(\omega \delta t) \nonumber\\
                & + \frac{\sqrt 2\gamma n_i}{\omega}\left( 1-\rm{cos}(\omega \delta t)\right)
\end{align}
and 
\begin{align}
 p_i(t + \delta t) = &  -x_i(t) {\rm{sin}}(\omega \delta t) + p_i(t){\rm{cos}}(\omega \delta t) \nonumber\\
                & + \frac{\sqrt 2\gamma n_i}{\omega}{\rm{sin}}(\omega \delta t).
\end{align}
In our article, we implemented the above equations incorrectly; we used the wrong signs for the second and first terms of the equations for $x_i(t+\delta t)$ and $p_i(t+\delta t)$, respectively, on the right-hand side.


Figure \ref{fig:Q62T} of this Erratum should replace Fig. 2 of the paper. The upper panel of Fig.~\ref{fig:Q62T} shows the pumped charge during $50$ pump cycles at resonance for the same parameters considered in our article. In Fig.~\ref{fig:Q62T}(a), the quantized pumping breaks down earlier in comparison to our article. In Fig.~\ref{fig:Q62T}(b), the direction of pumping changes thrice in the same time limit in contrast to the published result, where it happens twice. This is, however, parameter dependent. Other choices of parameters lead to two changes in the direction of pumping. There is no critical change in Fig.~\ref{fig:Q62T}(c). Analyzing the current result, we arrive at the same conclusions as before about the negative charge pumping and the breakdown of pumping at resonance. The lower panel of Fig.~\ref{fig:Q62T}, which shows the number of phonons $n_{ph}(t)$, modifies accordingly.

Figures~\ref{fig:62Tenr} and~\ref{fig:62Tenr1} should replace Figs. 3 and 4 of the paper, respectively. Figure~\ref{fig:62Tenr} (Fig.~\ref{fig:62Tenr1}) explains the underlying physics of the negative charge pumping (breakdown of pumping) in terms of the effective pumping path [Fig.~\ref{fig:62Tenr} (a-c) (Fig.~\ref{fig:62Tenr1} (a-c))], the energy gap at the center of the instantaneous spectrum [Fig.~\ref{fig:62Tenr}(e) (Fig.~\ref{fig:62Tenr1}(e)], and the occupancy of the lower and upper half of the instantaneous spectrum [Fig.~\ref{fig:62Tenr}(f) (Fig.~\ref{fig:62Tenr1}(f)]. Even though the figure changes quantitatively, the qualitative physics remains the same, and we can explain the physics the same way as in the article. 

In the corrected results, for Fig.~\ref{fig:62Tenr1} of the Erratum, the pumping path crosses the origin late compared to Fig. 4 of our paper, where it crosses the origin within the first two cycles. This leads to differences in $E_g$, $n_{up}$ and $n_{down}$ also. In conclusion, similar to the discussion in our paper, the non-adiabatic dynamics cause the breakdown of the topological charge pumping.

Fig.~\ref{fig:Q90T} of the Erratum, which should replace Fig. 5 of the paper, shows the same quantities where the pumping period is away from resonance. Here also, the approximate features remain the same as in the published results. The existence of a quantized pump at finite electron-phonon coupling [finite $\gamma$, Fig.~\ref{fig:Q90T}(a)] and breakdown at higher $\gamma$ [Fig.~\ref{fig:Q90T}(c)] can be seen in the upper panel. The behavior of $n_{ph}$ [Fig.~\ref{fig:Q90T}(d-f)] is also similar to the previous result. Here we have considered the time period of pumping as $T=90/J$ instead of $T=50/J$, which yields qualitatively the same behavior as in the original data. Also, we consider $\gamma = 0.2J,~0.5J$, and $1.0J$, respectively for Figs.~\ref{fig:Q90T}(a-c), which differs from the parameters of Fig. 5 our paper. However, at this new value of $T=90/J$, the updated values of $\gamma = 0.2J,~0.5J$, and $1.0J$ target the same parameter regimes of quantized pumping, the intermediate region, and breakdown of pumping, respectively, as in the paper.

The quantized pumping at finite $\gamma$ away from the resonance is explained with the help of an effective pumping path in Fig.~\ref{fig:62Tenr2}, along with other quantities. Figure~\ref{fig:62Tenr2} replaces Fig. 6 of the paper. The findings are the same as before; the stable trajectory of the effective pumping path around the origin protects the quantization. The actual path is different from the previous result, which is not important for topological protection.

Figure~\ref{fig:gap}, which should replace Fig. 7 of the paper, signifies the same physics as before. The gap closing of the instantaneous spectrum complements the breakdown of quantized pumping. The only difference is that we do not see clear negative charge pumping in this parameter regime ($T=90/J$), which is parameter-dependent.

The stability diagram, shown in Fig.~\ref{fig:pd}, also looks similar and shows all the features as before (replacing Fig. 8 of the paper); the parameter regime of quantized pumping, resonance, and reentrant quantization of pumping.

\begin{figure*}
\begin{center}
\includegraphics[width=1.\linewidth]{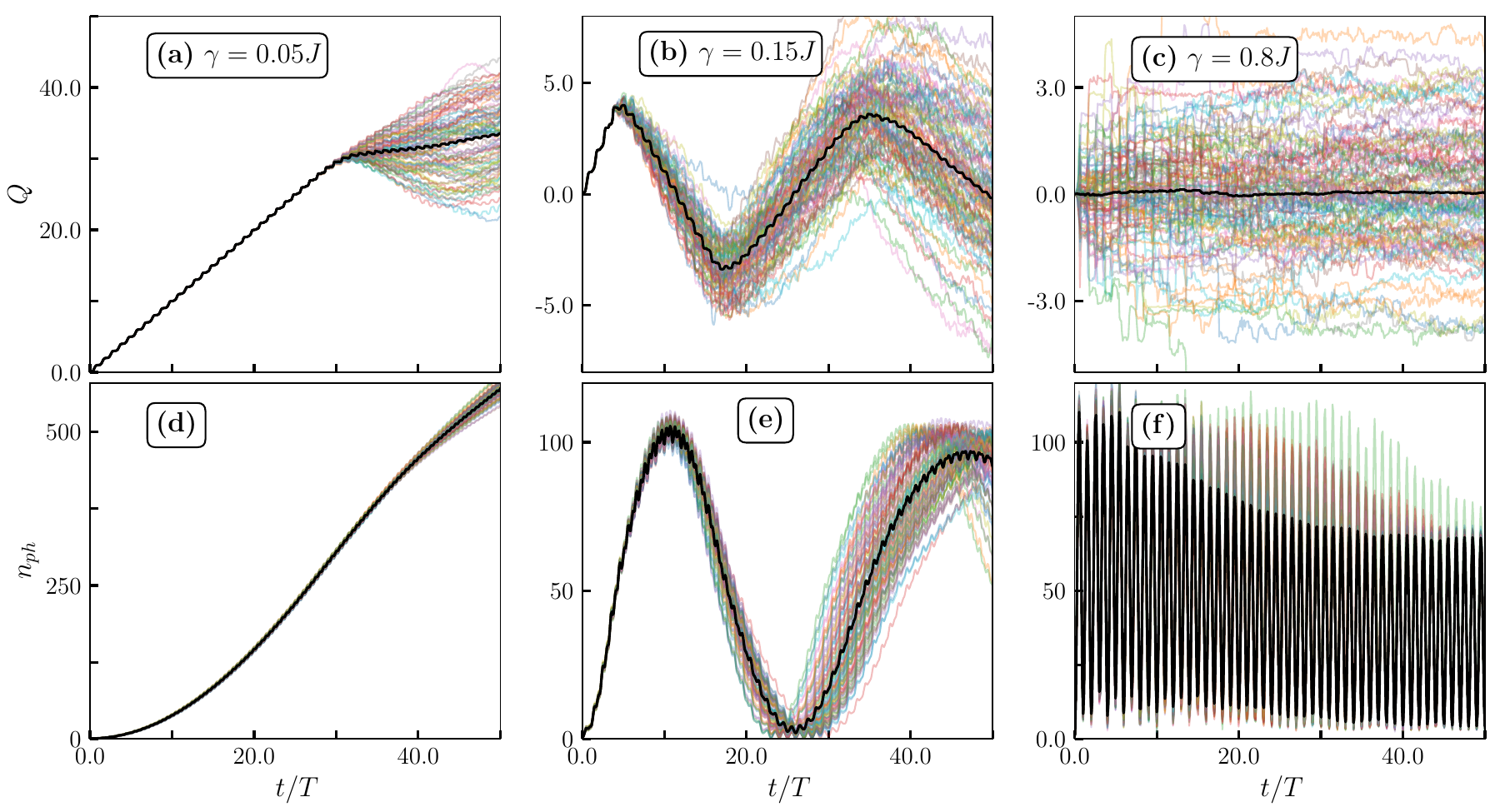}
 \end{center}
\vspace*{-0.7cm}
\caption{Observables obtained in the time evolution in 50 pump cycles at resonance for $N_{traj} = 100$ and $\omega = 0.1J$ with $T=2\pi/\omega$. The $Q(t)$ for all the trajectories (lines with transparent colours) and trajectory average $Q(t)$ (black line) is shown in (a-c) for $\gamma = 0.05J,~0.15J$, and $0.8J$, respectively. Similar to (a-c), we plot the phonon density $n_{ph}$ in (d-f), respectively.}
\label{fig:Q62T}
\vspace*{-0.17cm}
\end{figure*}


There are two minor differences. The total pumped charge goes to zero more gradually after the breakdown of quantization which can be observed from the gradual change in the shade of red in Fig.~\ref{fig:pd}(a) (away from resonance). Moreover, the breakdown regime of quantization moves slightly.

According to the discussion above, the article needs to be changed in the following places:



The last line of Sec. IIIA should be modified from "The counterclockwise winding ... is finite at later times." to "The anti-clockwise winding ... $E_g(t)$ is finite in this time window."

The second line in Sec. IIIB needs to be modified from "Here, the phonon coupling
drastically perturbs the staggering nature of the potential, and the effective pumping path becomes flat along the $\bar{\delta}$ axis, which crosses the origin in every pump cycle [Figs.4(a) - 4(c)]." to "Here, the phonon coupling drastically perturbs the staggered nature of the potential, and the effective pumping path {eventually} becomes flat along the $\bar{\delta}$ axis which crosses the origin in every pump cycle [Fig.~4(c)]."

The fourth line in Sec. IIIB needs to be modified from "The vanishing $E_g(t)$, shown in Fig. 4(e), signifies a nonadiabatic dynamics under $\hat H_{el}(t)$ leading to a breakdown of pumping." to "The vanishing $E_g(t)$ {in later pump cycles}, shown in Fig. 4(e), signifies a nonadiabatic dynamics under $\hat H_{el}(t)$ leading to the breakdown of pumping."
    
At the end of Sec. IIIB, the line "The $n_{\rm{dn}} \sim n_{\rm{up}}$ shown in ... equally mixed in both the bands" should be replaced by "The $n_{\rm{dn}}$ and $n_{\rm{up}}$ shown in Fig. 4(f) also suggests the excitation of the upper band as a result of non-adiabatic dynamics."

Since we are showing the out-of-resonance case for a different parameter set in the new results, the parameter values must be changed in some places. In the fifth line of the first paragraph of Sec. IV, the parameters should be updated to "$\gamma = 0.2J,~0.5J$, and $1.0J$". In the first line of the next paragraph, the "$\gamma$ to $0.4J$" should be replaced by "$\gamma$ to $0.5J$". Also, the last line of the same paragraph should be replaced with "Further increasing $\gamma = 1.0J$ ... $\gamma=0.5J$ case". In the third line of Sec. IVB, the parameter should be updated with $T=90/J$. Also, in the fifth line, there is a change of parameter from ($\gamma \approx 0.4J$) to ($\gamma \approx 0.5J$). The last line of Sec. IVB should be removed since, at this parameter set, we do not see negative charge pumping, which is a parameter dependent phenomenon.

In the first line of Sec. V, the parameter should be updated to ($T=\frac{2\pi}{0.1J}$ and $90/J$).

\phantom{10cm}

\clearpage

\begin{figure}[H]
\begin{center}
\includegraphics[width=1.\linewidth]{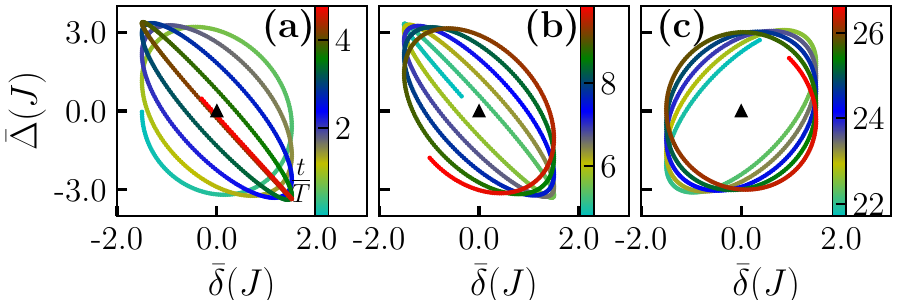}
\includegraphics[width=1.\linewidth]{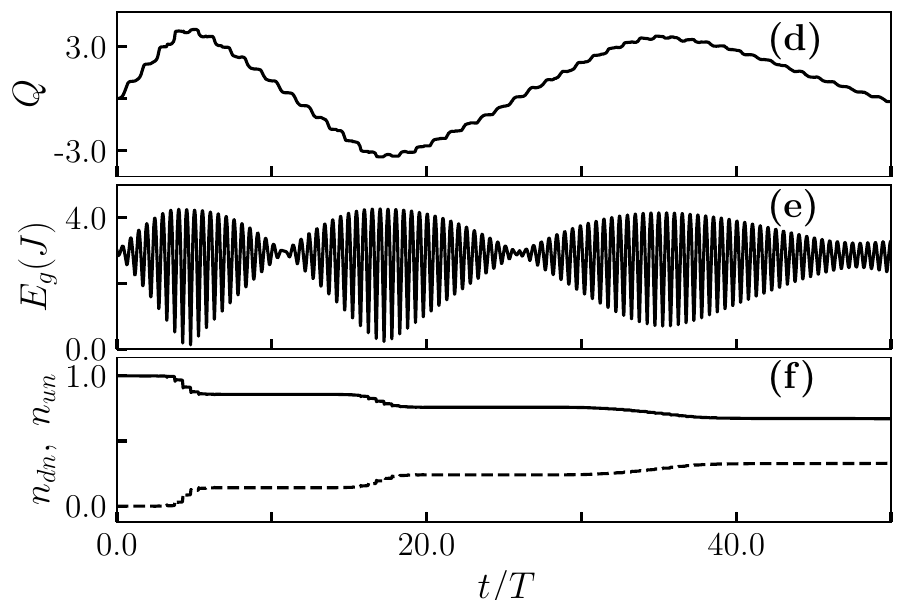}
\end{center}
\vspace*{-0.7cm}
\caption{Trajectory averaged quantities are displayed here at resonance for $\gamma=0.15J$, $\omega =0.1J$ and $T = 2\pi/\omega$ with $N_{traj} = 100$. The trajectory averaged effective pumping path is shown for three different time windows in (a), (b), and (c) corresponding to Fig.~\ref{fig:Q62T}(b), which is also shown in (d) for reference. The black triangles mark the origins where the gap should vanish. The direction of winding around the origin signifies the direction of pumping. The $E_g(t)$ is shown in (e) and $n_{\rm{dn}}(t)$ (solid line) and $n_{\rm{up}}(t)$ (dashed line) are plotted in (f).}
\label{fig:62Tenr}
\end{figure}

\begin{figure}[H]
\begin{center}
\includegraphics[width=1.\linewidth]{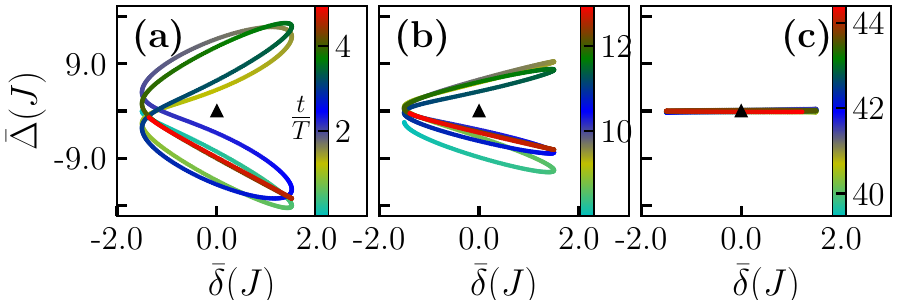}
\includegraphics[width=1.\linewidth]{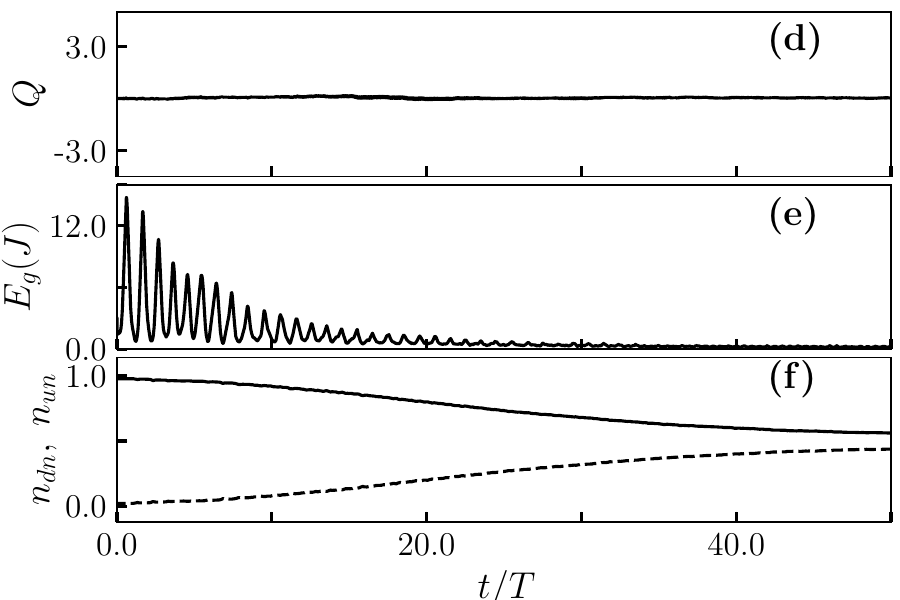}
\end{center}
\vspace*{-0.7cm}
\caption{Explanation of the breakdown of pumping at resonance for $\gamma=0.8J$, $\omega =0.1J$ and $T = 2\pi/\omega$. Different quantities are plotted after averaging over $100$ trajectories. The effective pumping path is shown for three different time windows in (a), (b), and (c) corresponding to Fig.~\ref{fig:Q62T}(c), which is also shown in (d) for reference. The black triangles mark the origins where the gap should vanish. The frequent crossing of origin by the effective pumping path in later pump cycles signifies the breakdown of TCP. The $E_g(t)$ is shown in (e) and $n_{\rm{dn}}(t)$ (solid line) and $n_{\rm{up}}(t)$ (dashed line) are plotted in (f).}
\label{fig:62Tenr1}
\vspace{-2cm}
\end{figure}

\begin{figure*}
\includegraphics[width=1\linewidth]{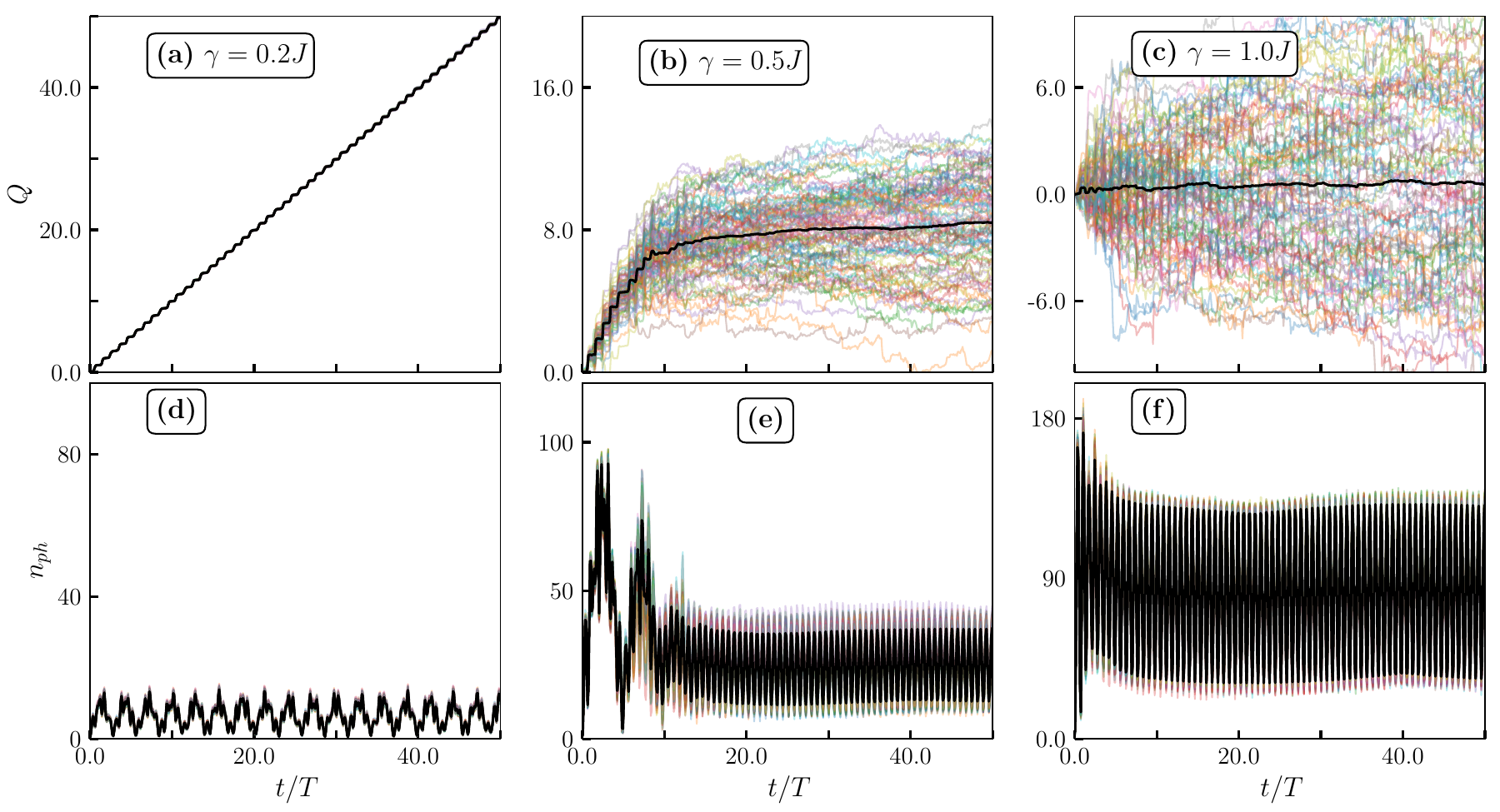}
\label{fig:Q90T}
\caption{Observables obtained in the time evolution in 50 pump cycles away from resonance for $N_{traj} = 100$ and $\omega = 0.1J$ with $T=90/J$. The $Q(t)$ for all the trajectories (lines with transparent colours) and trajectory average $Q(t)$ (black line) is shown in (a-c) for $\gamma = 0.2J,~0.5J$, and $1.0J$, respectively. Similar to (a-c), we plot the phonon density $n_{ph}$ in (d-f), respectively.}
\label{fig:Q90T}
\end{figure*}


\begin{figure}[H]
\begin{center}
\includegraphics[width=1.\linewidth]{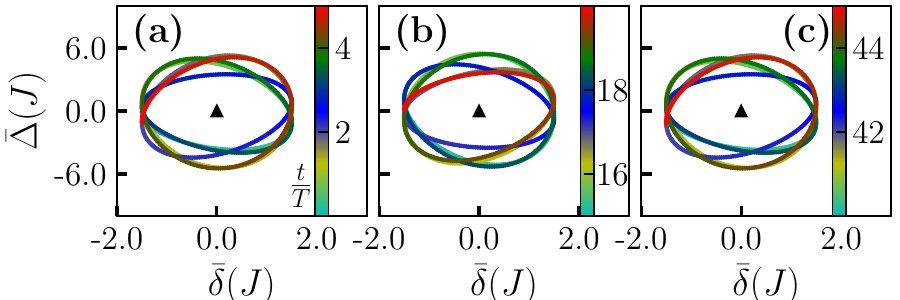}
\includegraphics[width=1.\linewidth]{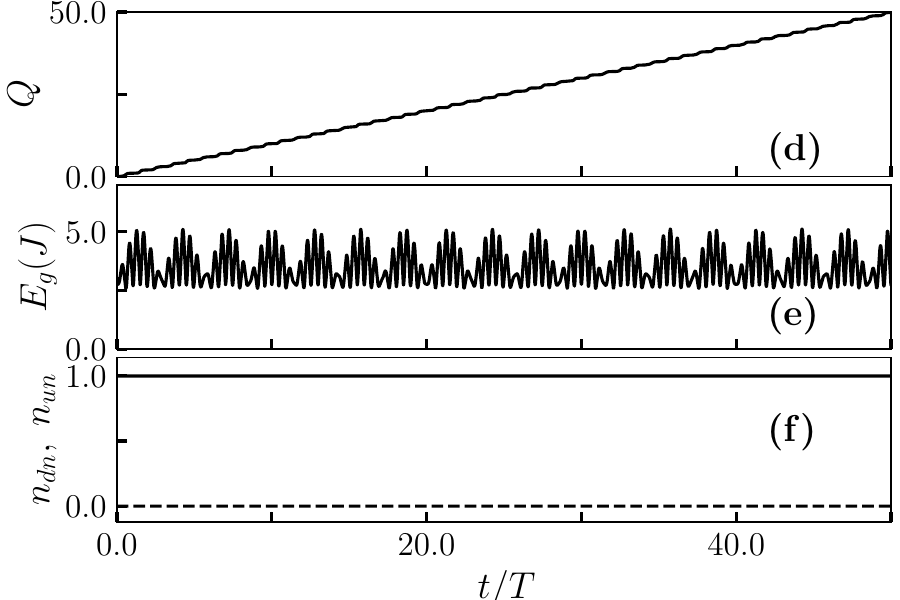}
\end{center}
\caption{Different quantities are displayed here after averaging over trajectories with $N_{traj} = 100$ to analyze the quantized pumping at finite $\gamma=0.2J$ away from resonance where $\omega =0.1J$ and $T = 90/J$. The effective pumping path is shown for three different time windows in (a), (b), and (c), corresponding to Fig.~\ref{fig:Q90T}(a), which is also shown in (d) for reference. The black triangles mark the origins where the $E_{g}$ should vanish. The $E_g(t)$ is shown in (e) and $n_{\rm{dn}}(t)$ (solid line) and $n_{\rm{up}}(t)$ (dashed line) are plotted in (f).}
\label{fig:62Tenr2}
\end{figure}

\begin{figure}[H]
\begin{center}
\includegraphics[width=1.\linewidth]{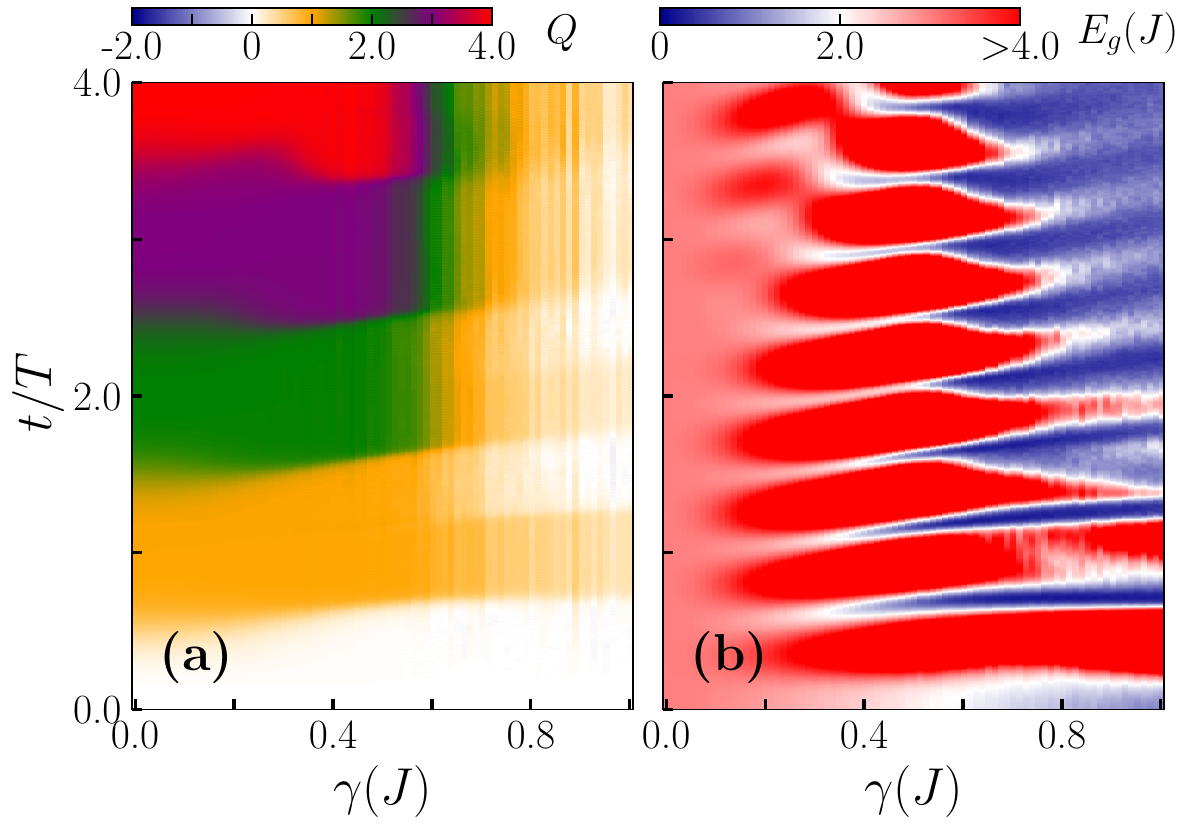}
 \end{center}
\caption{(a) The trajectory averaged $Q(t)$ and (b) $E_{gap}(t)$ are shown away from resonance with $T=90/J$, $\omega = 0.1J$ and $N_{traj} = 200$. The breakdown of quantized pumping near {$\gamma \sim 0.5J$} is portrayed}
\label{fig:gap}
\end{figure}

\begin{figure}[H]
\begin{center}
\includegraphics[width=1.\linewidth]{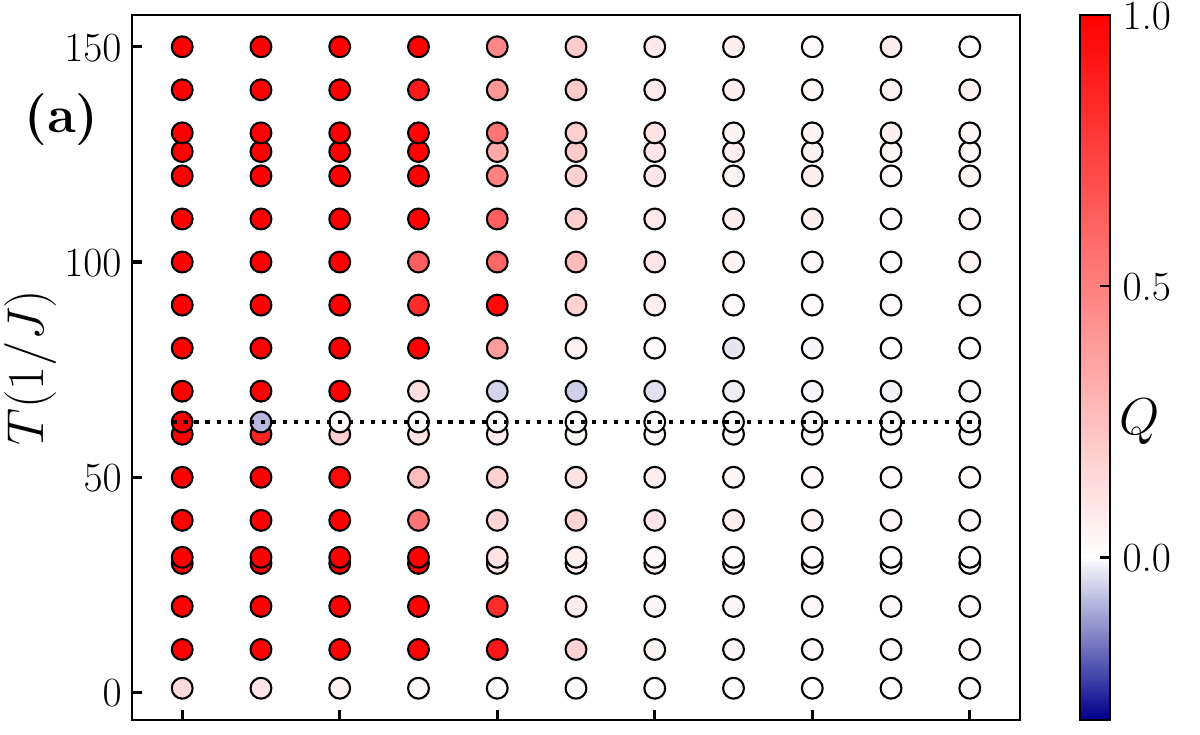}
\includegraphics[width=1.\linewidth]{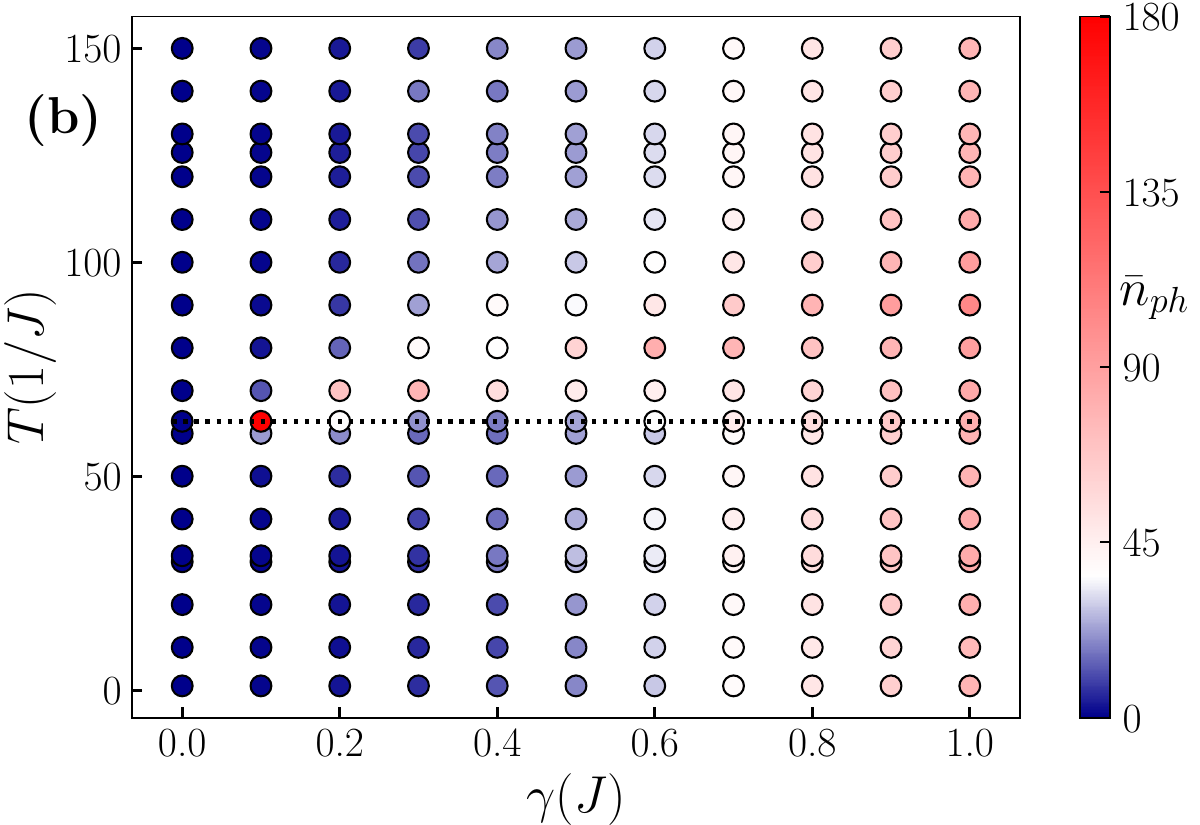}
 \end{center}
\caption{The figure represent a stability diagram in $T$ vs $\gamma$ plane depending on the trajectory averaged (a) $Q$ in a cycle and (b) $\bar{n}_{ph}$. Here we consider $N_{traj} = 100$ and average the $Q$ and $\bar{n}_{ph}$ over fifty pump cycles. The dashed lines represent the time period of phonons ($\frac{2\pi}{\omega}$).}
\label{fig:pd}
\end{figure}

\clearpage

\end{document}